\documentclass[journal, 12pt, draftclsnofoot, onecolumn]{IEEEtran}
\usepackage[acronym,shortcuts,nonumberlist,hyperfirst=false,nopostdot,nogroupskip,nomain]{glossaries}
\usepackage{cite}
\usepackage[pdftex]{graphicx}
\graphicspath{{./figures}}
\DeclareGraphicsExtensions{.pdf,.jpeg,.png} 
\usepackage{tikz}
\usetikzlibrary{arrows, automata, shapes.geometric, positioning, decorations.pathreplacing, calc, matrix}
\usepackage{amsmath,bm}
\usepackage{amssymb}
\usepackage{comment}
\usepackage{algorithmic}
\usepackage{algorithm}
\usepackage{caption}
\usepackage{subcaption}
\usepackage{array}
\usepackage{mathtools}

\DeclareMathSymbol{\shortminus}{\mathbin}{AMSa}{"39}
\newtheorem{defn}{Definition}

\newcommand{\Ia}{\mathcal{I}_{\mathrm{a}}}
\newcommand{\Ka}{K_{\mathrm{a}}}
\newcommand{\Ktot}{K}
\newcommand{\I}{\mathcal{I}}
\newcommand{\ASD}{\mathbf{A}^{(\mathrm{U})}}
\newcommand{\ACD}{\mathbf{A}^{(\mathrm{S})}}

\newcommand{\XCD}{\mathbf{X}^{(\mathrm{S})}}
\newcommand{\XSD}{\mathbf{X}^{(\mathrm{U})}}

\newcommand{\ntot}{N_{\mathrm{tot}}}
\newcommand{\noise}{\mathbf{w}}
\newcommand{\bignoise}{\mathbf{W}}

\newcommand{\blck}{N}  
\newcommand{\info}{b}

\newcommand{\mw}{M}

\newignoredglossary{ignore}

\newacronym{AMP}{AMP}{approximate message passing}
\newacronym{MMV-AMP}{MMV-AMP}{multiple measurement vector approximate message passing}
\newacronym{MMV-AMP1}{MMV-AMP}{multiple measurement vector AMP}
\newacronym{SISO}{SISO}{soft-input-soft-output}

\newacronym{SNR}{SNR}{signal-to-noise-ratio}

\begin{document}
%
\title{Constant Weight Codes with Gabor Dictionaries and Bayesian Decoding for Massive Random Access}
%
\author{Patrick Agostini, Zoran Utkovski, Alexis Decurninge, Maxime Guillaud, \\and S\l awomir Sta\'{n}czak
\thanks{Patrick Agostini, Zoran Utkovski and S\l awomir Sta\'{n}czak are with the Department of Wireless Communications and Networks, Fraunhofer Heinrich-Hertz  Institute in Berlin, Germany.  Alexis Decurninge and Maxime Guillaud are with  the Mathematical and Algorithmic Sciences Laboratory, Paris Research Center, Huawei Technologies France. Corresponding author: Zoran Utkovski (zoran.utkovski@hhi.fraunhofer.de).}}
%
\maketitle
%
\vspace{-20pt}
\begin{abstract}
    This paper considers a general framework for massive random access based on sparse superposition coding. We provide guidelines for the code design and propose the use of constant-weight codes in combination with a dictionary design based on Gabor frames. The decoder applies an extension of approximate message passing (AMP) by iteratively exchanging soft information between an AMP module that accounts for the dictionary structure, and a second inference module that utilizes the structure of the involved constant-weight code. We apply the encoding structure to (i) the unsourced random access setting, where all users employ a common dictionary, and (ii) to the "sourced" random access setting with user-specific dictionaries. When applied to a fading scenario, the communication scheme essentially operates non-coherently, as channel state information is required neither at the transmitter nor at the receiver. We observe that in regimes of practical interest, the proposed scheme compares favorably with state-of-the art schemes, in terms of the (per-user) energy-per-bit requirement,  as well as the number of active users that can be simultaneously accommodated in the system. Importantly, this is achieved with a considerably smaller size of the transmitted codewords, potentially yielding lower latency and  bandwidth occupancy, as well as lower implementation complexity. 
\end{abstract}
%
%
%
\IEEEpeerreviewmaketitle
%
%
\vspace{-17.5pt}
\section{Introduction}
\IEEEPARstart{T}{he} growing interest in Internet-of-Things (IoT) applications has put massive Machine Type Communications (mMTC) at the focus of the wireless communications research for Post-5G Networks. mMTC services are characterized by the presence of a potentially massive number of terminals that transmit short packets in a sporadic fashion, with potential applications in various domains, ranging from industry and smart city to logistics and healthcare.

The design of mMTC communication protocols is challenged by the inadequacy of the conventional Multiple Access Channel (MAC) model that has provided the theoretical background for the study of uplink transmission strategies. Unlike the standard MAC model, mMTC systems are typically characterized by small payloads, uncoordinated access and sporadic user activity, with the number of active users possibly exceeding the overall  message blocklength. While massive random access in the context of IoT has attracted considerable attention, a unifying framework that structures the different approaches has been elusive. The situation is made further more complex due to the different modelling assumptions that come with different approaches.\smallskip

\noindent \textit{\textbf{The Context for Massive Random Access:}} From an information-theoretic perspective, random access is related to the conventional MAC model. Early information-theoretic studies on the MAC (see e.g.~\cite{Ahlswede1973}), treat the MAC in the so-called ergodic regime where the fundamental limits are studied in the asymptotic limit of infinite coding block-length. In particular, the MAC capacity region is computed assuming that the set of transmitting users is typically small and known in advance, and the users are coordinated in terms of the protocol for channel access. 
Extending over the MAC model, random user activation has been integrated in information-theoretic models by way of partially active users (“T-out-of-N MAC”)~\cite{Mathys1990, bar-david93}. Different to the conventional MAC and the extensions therein, in massive IoT the packets are usually short and the control information, such as the user address, has a size that is comparable to the data size. This was addressed in the context of the "many-access" channel model (MnAC)~\cite{chen14, Chen2017}, which accommodates random activation and allows the number of users to increase proportionally to the blocklength. This is in contrast to the standard large-system analysis of multiuser systems in which the blocklength goes to infinity before the number of users is made arbitrarily large~\cite{shamai97}. Recently proposed random access protocols based on the concept of unsourced random access (U-RA)~\cite{polyanskiy17} address the transmission of short packets from a different perspective. Within this framework, the users employ the same codebook and collisions are interpreted as the event where multiple users transmit the same codewords. As a consequence of the shared codebook, the problem of user identification is separated from the actual data transmission, and the decoder only declares which messages were transmitted, without associating the messages to the transmitting users. 

\noindent\textit{\textbf{Sparse Regression Codes (SPARCs):}} SPARCs, also known as Sparse Superposition Codes, have been introduced by Joseph and Barron \cite{joseph12}, \cite{joseph14} for the memoryless (point-to-point) AWGN. A sparse superposition code is defined in terms of a design matrix $\mathbf{A}$ of dimension $\blck\times ML$ with i.i.d. $\mathcal{CN}(0, 1)$ entries. 
Here $\blck$ is the block length while $M$ and $L$ are integers that define the code rate. One can think of the matrix $\mathbf{A}$ as composed of $L$ sections with $M$ columns each. Each codeword is a superposition of $L$ columns, with one column from each section. 
An efficient decoding algorithm for sparse superposition codes called "adaptive successive decoding" has been proposed in \cite{joseph12}. 
Subsequently, a soft-decision iterative decoder was proposed in \cite{barron12}, with theoretical guarantees similar to the earlier decoder in \cite{joseph12} but improved empirical performance for finite block lengths. In \cite{rush17}, the authors proposed an AMP-based decoder, with probability of decoding error going to zero with growing block length $N$ for all fixed rates $R<C$. 

In the multiuser scenario, the $L$ sections may be interpreted as user-specific sets of sequences $\mathbf{A}_l=[\mathbf{a}_{1l},\dots,\mathbf{a}_{Ml}]\in\mathbb{C}^{\blck\times M}$ of cardinality M, assigned to each user $l\in[L]$. A variant of this approach has been used  for a joint device detection and data transmission in an mMTC setup~\cite{Larsson_1}, \cite{Liu2018}. Non-coherent transmission was achieved by encoding the information into the choice of the sequence sent by the device, mapping $\info$ bits onto $2^{\info}$ possible pilots. The SPARCs framework has been recently  applied to the massive unsourced random access setting by Fengler et al.~\cite{fengler2019ISIT}, \cite{Fengler2021}, demonstrating state-of-the art performance on the real Gaussian MAC \textcolor{black}{and massive MIMO fading channels in \cite{nonBayesian}. More recently, the SPARCs framework in combination with a cooperative activity detection framework for sixth-generation (6G) cell-free has been introduced in \cite{sourcedUnsourcedMRA} for cooperative unsourced random access.}
\vspace{-10pt}
\subsection{Main Contributions}
\textcolor{black}{In this paper, we consider a general framework for massive random access based on sparse superposition coding, according to which  the users convey information by linearly combining sequences from a predefined dictionary based on an appropriate error-correction code.} 
\textcolor{black}{The proposed transmission scheme comprises three main ingredients: (i) sparse-superposition coding based on constant-weight codes, (ii) dictionary design based on Gabor frames and (iii) Bayesian decoding based on an adaptation of \textcolor{black}{\ac{AMP} \cite{ranganDamping}, respectively  \ac{MMV-AMP1} for the multi-antenna scenarios \cite{kim2011belief}.} The decoding simultaneously accounts for the dictionary structure, as well as for the structure of the underlying code.} \smallskip


\noindent \textit{\textbf{Motivation:}} To motivate the approach, consider an AWGN scenario with an active set $\mathcal{I}_a$ of users sharing the same dictionary matrix $\mathbf{A}$. Upon transmission, an active user $i\in \mathcal{I}_a$ applies the  dictionary matrix $\mathbf{A}$ to map a (sparse) binary codeword $\mathbf{c}_i$ to a transmit signal $\mathbf{s}_i=\mathbf{A}\mathbf{c}_i$. The receiver observes the linear mixture\smallskip
\begin{align}
    \mathbf{y}=\mathbf{A}\mathbf{c}+\noise,
    \label{eq:example}
\end{align}
where $\noise$ is an AWGN noise vector of corresponding dimension, and $\mathbf{c}$ is the superposition (over the reals) of the binary codewords transmitted by the active users, $\mathbf{c}=\sum_{i\in \mathcal{I}_a}\mathbf{c}_i$.\smallskip

Now, consider the (simplified) decoding setup consisting of two decoding steps: i) in the first step the decoder employs a compressive sensing-based reconstruction  algorithm that estimates the support of the vector $\mathbf{c}$, $\mathrm{supp}(\mathbf{c})$, according to \eqref{eq:example}; ii) in the second step the decoder retrieves the individual messages of the active users from $\mathrm{supp}(\mathbf{c})$. A key observation is that this decoding process yields an \textit{effective channel} that can be viewed as a concatenation of an \textit{inner channel} \smallskip
\begin{equation}
\mathrm{supp}(\mathbf{c})\rightarrow \mathbf{A}\mathbf{c}+\noise, 
\end{equation}
and an \textit{outer channel} (output of the first decoding step), which is in effect a (noisy) binary input \textit{OR-MAC channel} \smallskip
\begin{equation}
    \bigvee\limits_{i\in\mathcal{I}_{\mathrm{a}}} \mathbf{c}_{i}\rightarrow \mathrm{supp}(\mathbf{c}),
\end{equation}
where $\vee$ denotes the binary OR operation (performed component-wise). 

Hence, the role of the second decoding step, i.e. the \textit{outer decoder}, is to output the individual binary codewords $\{\mathbf{c}_i\}_{i\in\mathcal{I}_a}$ from their OR-superposition $\bigvee_{i\in\mathcal{I}_{\mathrm{a}}}\mathbf{c}_{i}$. Consequently, the aim of the \textit{outer encoder} is to ensure that the binary codewords of the active users can be retrieved from their OR superposition with high probability. This aspect is closely related to the concept of \textit{uniquely decipherable codes}~\cite{Kautz}. Formally, a uniquely decipherable code of order $p$ has the property that every OR superposition of up to $p$ different codewords is distinct from every other superposition of $p$ or fewer codewords. As we will see in Section~\ref{sec:Code_Design}, a somewhat more special is the concept of a \textit{disjunctive code}\cite{AbdulJabbar88}, which besides being uniquely decipherable, also fulfils the so-called \textit{zero-false-drop} condition. In this context, the rationale  behind applying \textit{constant-weight} (CW) codes stems form the observation that disjunctive codes of a required order may be obtained from CW codes with an appropriate parametrization. 

Accordingly, the role of the \textit{inner encoder} in the described setup would be to ensure that the \textit{inner decoder} can reliably estimate $\mathrm{supp}(\mathbf{c})$ from the observation $\mathbf{y}$ in the first decoding step. For this purpose, we advocate the use of Gabor frames to design the dictionary matrix $\mathbf{A}$, due to their excellent coherence properties, reduced storage and encoding/decoding requirements.

\noindent\textit{\textbf{Summary of the contributions:}}
\begin{itemize}
	\item We provide guidelines for the code design for massive random access and, in particular, propose the use of \textit{constant-weight} codes in combination with (approximate) Bayesian message passing decoding. We formulate a general encoding structure that we apply to the (U-RA) setting where all users employ a common (shared) dictionary, as well as to the "sourced random access" (S-RA)  setting where the individual users are assigned separate (i.e. user-specific) dictionaries, as in, e.g., the many-access channel model~\cite{chen14},~\cite{Chen2017}. \smallskip 

	\item We address both a Gaussian channel model, as in e.g. \cite{ordentlich17} and \cite{Fengler2021}, and a \textit{non-coherent} block-fading scenario where the users neither employ precoding nor dedicate resources (transmit training signals) to enable channel estimation. This is of particular relevance for the massive access setup with sporadic transmissions of short messages, for which we argue that a meaningful performance analysis should drop the assumption of a priori CSI at both the receiver and at the transmitter (see, e.g., \cite{durisi16}, \cite{devassy15} for related discussions for the MAC).\smallskip

	\item For efficient receiver processing we use an adaptation of the AMP algorithm for approximate Bayesian inference~\cite{kim2011belief}. The inference algorithm operates by \textcolor{black}{exchanging information} between two modules: the first module carries out standard AMP and accounts for the dictionary structure, but ignores the dependencies imposed by the CW code that prescribes how the information messages are mapped to the linear combinations of dictionary elements; the second inference module refines the output of the AMP module by handling the dependencies coming from the structure of the applied code. In this context, the applied code for sequence selection can be interpreted as an outer code that provides error correction for the "effective" communication channel at the output of the AMP module.\smallskip 

	\item We propose a dictionary design based on a finite \textit{Gabor frame} obtained from the Alltop seed vector, which shows excellent coherence properties.  Gabor frames are completely specified by the seed vector, and multiplications with Gabor frames can be efficiently carried out using FFT, resulting in reduced encoding/decoding complexity and memory requirements. We note that the use of Gabor dictionaries in combination with CW codes yields a form of coded modulation that is tailored to the multiple access scenario with short messages (see, e.g., \cite{decurninge2021tensor} for an example of tensor-based modulation for unsourced random access.)
	\smallskip
	
	\item The numerical simulations indicate performance advantages over state-of-the art schemes in terms of the energy-per-bit required by each user to meet a target error probability, and the number of active users that can be simultaneously accommodated in the system. In parallel, the transmission scheme operates with comparably short transmit codeword sizes, yielding lower latency and  bandwidth occupancy, as well as lower  complexity of implementation.
\end{itemize}
The remaining of the paper is organized as follows. In Section~\ref{sec:system_model} we introduce the system model and describe the encoding procedure for the two studied scenarios. Details of the code design are given in Section~\ref{sec:Code_Design}. A high-level description of the receiver architecture and the decoding procedures is provided in Section~\ref{sec:decoding}, while the full details are given in Appendix~\ref{sec:appendix}-\ref{sec:appendix_outer_decoder}. Section~\ref{sec:dictionary_design} provides details of the proposed dictionary design. The performance of the proposed transmission schemes is assessed via numerical simulations in Section~\ref{sec:results}. Section~\ref{sec:conclusion} concludes the paper.\smallskip

\noindent \textit{\textbf{Notation:}} Unless specified otherwise, we use lower- and upper-case bold letters to denote vectors and matrices, respectively. Upper-case calligraphic letters denote sets. $\mathbf{I}$ denotes the identity matrix of corresponding dimension. We use $\mathbf{x}\sim\mathcal{CN}(\mathbf{m},\mathbf{R})$ to denote that the random vector $\mathbf{x}$ follows a circularly-symmetric complex Gaussian distribution with mean $\mathbf{m}$ and covariance matrix $\mathbf{R}$. We use $\{\cdot\}^{\mathrm{T}}$ and $\{\cdot\}^{\mathrm{H}}$ for the transpose and Hermitian operators, respectively. By $[N]:=\{1,\ldots,N\}$ we denote the counting set of cardinality $N$.
\vspace{-6pt}
\section{System Model and Encoding} \label{sec:system_model}
\subsection{System Model}
Consider a multiple access channel where communication is performed over blocks of $N$ channel uses. We consider a pool of $K$ possible single-antenna users, indexed by the set $\I:=\{1,\dots,K\}$, that can transmit in every block. The activity of a user $i\in\mathcal{I}$ is assumed to be unknown to the receiver and is captured by the binary random variable $\mathrm{\lambda}_i\in\{0,1\}$, with $\lambda_i=1$ if the user $i$ belongs to the active set, $i\in\Ia$, and $\lambda_i=0$ otherwise. We denote by $\Ka$ the number of active users in a considered transmission block. 
When active, user $i\in\Ia$ selects a message $W_i\in[2^{\info}]$ of $\info$ bits from its transmit buffer and spreads it over its transmission block
by generating a transmit vector (transmit codeword) $\mathbf{s}_i\in\mathbb{C}^{\blck}$.
We further assume that all active users $i\in\Ia$ are subject to the same power constraint $P$ such that $(\forall i\in\Ia)$  $\mathbb{E}[{\Vert\mathbf{s}_i\Vert^2}]=PN$. 

Assuming a block-fading model with coherence time-frequency span no smaller than the block size $N$ and a single-antenna receiver, the received signal $\mathbf{y}\in\mathbb{C}^{\blck}$ reads
\begin{equation}
	\mathbf{y} = \sum_{i\in\mathcal{I}_{\mathrm{a}}}h_{i}\mathbf{s}_{i}+\noise, \label{eq:system_model}
\end{equation} 
where $h_{i}\in\mathbb{C}$ is the channel coefficient for the wireless channel between user $i$ and the receiver for the considered resource block. The entries of the noise vector $\noise$ are i.i.d. complex Gaussian with zero mean and variance $N_0$. With this, the per-user energy-per-bit is defined as $\frac{E_b}{N_0}\doteq\frac{PN}{b N_0}$. Importantly, we consider a \textit{ non-coherent} setting without a priori CSI at the transmitter and at the receiver, with the receiver knowing only the statistics of the fading process.
\vspace{-20pt}
\subsection{Encoding}
The encoding procedure is as follows. Each active user $i\in\Ia$ first encodes its message $W_i\in[2^{\info}]$ into a binary vector $\mathbf{c}_i\in\{0,1\}^{\mw}$. The binary vector is then mapped to a vector
\begin{equation}
    \mathbf{s}_i=\mathbf{A}_i\mathbf{c}_i\label{eq:dictionary_mapping}
\end{equation}
of length $\blck$ by a matrix $\mathbf{A}_i\in\mathbb{C}^{\blck\times\mw}$ and transmitted over $\blck$ channel resources. The product is properly scaled such as to fulfil the average power constraint $\mathbb{E}[\Vert\mathbf{s}_i\Vert^2]=PN$. The matrices $\{\mathbf{A}_i\}_{i\in\Ia}$ can be interpreted as user dictionaries whose columns are (complex-valued) sequences of dimension $\blck$. The binary vectors $\{\mathbf{c}_i\}_{i\in\Ia}$ are mapped on the resource block via \eqref{eq:dictionary_mapping}, which determines the sequences from the dictionary $\mathbf{A}_i$ that are combined to produce the transmit vector $\mathbf{s}_i$. Depending on the dictionaries $\mathbf{A}_i$, we will distinguish between the following approaches. 

\subsubsection{Shared Dictionary}\label{sec:common_dictionary}
According to this approach, all users employ a shared (i.e. a common) dictionary $\mathbf{A}_1=\ldots=\mathbf{A}_K=\ACD$, to map their codewords $\mathbf{c}_i$ on a resource block of size $\blck$. Following \eqref{eq:system_model}, the receive signal over a resource block reads 
\begin{align}
    \mathbf{y} &=\ACD\sum\limits_{i\in\mathcal{I}_{\mathrm{a}}} h_i\mathbf{c}_i+\noise=\ACD\mathbf{x}+\noise,\label{eq:CD}
\end{align}
where we have defined $\mathbf{x}=\sum_{i\in\mathcal{I}_{\mathrm{a}}} h_i\mathbf{c}_i$. This approach is well suited to the U-RA model in \cite{polyanskiy17}, \cite{ordentlich17}, where the user identification problem and the data transmission problem are decoupled. In that case, the role of the decoder is to output a list $\mathcal{L}(y)=\{W_1,\ldots,W_J\}\subset[2^\info]^J$ of no more than $K_{\mathrm{a}}$ messages ($J\leq K_{\mathrm{a}})$ that should contain most messages that were transmitted by the active users, where the order in which the messages appear in the list is of no significance. In other words, the decoder only declares which messages were transmitted, without associating the messages to the transmitting users. With this, the decoder's error probability is defined as
\begin{equation}
    P_{\mathrm{e}}=\frac{1}{K_{\mathrm{a}}}\sum_{i\in\mathcal{I}_{\mathrm{a}}}\mathbb{P}\left(W_i\notin \mathcal{L} (\mathbf{y})\right).
    \label{eq:decoding_error_JC}
\end{equation}
We note that this formulation requires that the number of active users $K_{\mathrm{a}}$ is \textit{known} to the decoder. As a consequence, the number of users $K$ that can transmit on the block of $\blck$ channel uses does not affect the error probability, and can thus be left out of the model (i.e. can be set to $\infty$). 

\subsubsection{User-specific Dictionaries} \label{sec:separate_dictionaries}

In this case, the users employ user-specific (i.e. separate) dictionary matrices to map their binary codewords $\mathbf{c}_i$ on the channel resources. Therefore, considering \eqref{eq:system_model}, the signal received over the resource block of length $\blck$ is given by
\begin{align}
    \mathbf{y} &= \sum\limits_{i\in\I}\lambda_i h_{i}\ASD_{i}\mathbf{c}_{i}+\noise. \label{eq:SD}
\end{align}
Using $\ASD=[\ASD_{1},\ldots,\ASD_{K}]\in\mathbb{C}^{\blck\times \Ktot\mw}$ to denote a matrix that concatenates all $i\in\I$ dictionary matrices, \eqref{eq:SD} can be written in a more compact form as
\begin{equation} 
    \mathbf{y}=\ASD\mathbf{x}+\noise, \label{eq:separate_dictionaries}
\end{equation} 
where we define $\mathbf{x}=[\lambda_1 h_{1}\mathbf{c}_1;\ldots;\lambda_K h_{K}\mathbf{c}_K]\in \mathbb{C}^{\Ktot\mw}$. Different from the U-RA model, the use of user-specific dictionaries allows for user identification (hence it is suitable to S-RA). We  note that in this setting the number of system users is finite and is thus a parameter in the model.\smallskip 

\noindent \textit{\textbf{Extension to a multi-antenna setting:}} The extension of \eqref{eq:system_model} to the scenario in which the receiver is equipped with $T$ antennas (single-input-multiple-output (SIMO) scenario) is straightforward. In that case, the received signal $\mathbf{Y}\in \mathbb{C}^{N\times T}$ is given by  
\begin{equation}
	\mathbf{Y} = \sum_{i\in\mathcal{I}_{\mathrm{a}}}\mathbf{s}_{i}\mathbf{h}^{\mathrm{T}}_{i}+\bignoise, \label{eq:system_model_MIMO}
\end{equation}
where $\mathbf{h}_{i}\in\mathbb{C}^{T}$ denotes the wireless channel vector of user $i\in\mathcal{I}_{\mathrm{a}}$, and $\bignoise\in\mathbb{C}^{N\times T}$ is the additive receiver noise matrix, whose entries are i.i.d. complex Gaussian with zero mean and variance $N_0$. As a consequence, the extension of \eqref{eq:CD} to the multi-antenna setting is given by $\mathbf{Y}=\ACD\mathbf{X}^{(\mathrm{S})}+\bignoise$, 
with $\mathbf{X}^{(\mathrm{S})}=\sum_{i\in\mathcal{I}_{\mathrm{a}}} \mathbf{c}_i\mathbf{h}_i^{\mathrm{T}}\in \mathbb{C}^{M\times T}$. Similarly, the extension of \eqref{eq:separate_dictionaries} can be given by $\mathbf{Y}=\ASD\mathbf{X}^{(\mathrm{U})}+\bignoise$, 
with $\mathbf{X}^{(\mathrm{U})}=[\lambda_1 \mathbf{c}_1\mathbf{h}_1^{\mathrm{T}},\ldots,\lambda_K \mathbf{c}_K\mathbf{h}_K^{\mathrm{T}}]\in \mathbb{C}^{\Ktot\mw\times T}$. For simplicity, in the description that follows we will only consider the single-antenna setting. We note that details of the Bayesian inference procedure for the extended system model are provided in Appendix~\ref{sec:appendix}. Numerical results for the multi-antenna setting are provided in Section~\ref{sec:results} as part of our numerical evaluation.\smallskip

\noindent \textcolor{black}{\textit{\textbf{Extension to frequency selective fading channels:}} For frequency selective fading channels \eqref{eq:system_model} does not hold. Such channels are typically modelled as tapped-delay-line (TDL) filters with $P>1$ taps simulating the number of multipath components in the channel \cite{frequencySelective}, i.e., 
\begin{equation}
    \mathbf{y}=\sum_{i\in\Ia}\mathrm{diag}(\mathbf{s}_i)\mathbf{F}\tilde{\mathbf{h}}_i+\mathbf{w}=\sum_{i\in\Ia}\sum_{m\in[M]}c_{i,m}\sum_{p\in[P]}\tilde{h}_{i,p}\mathrm{diag}(\mathbf{a}_m)\boldsymbol{\omega}_p+\mathbf{w}\label{eq:freq_multitap}
\end{equation}
where $\tilde{\mathbf{h}}_i=[\tilde{h}_{i,0},\ldots,\tilde{h}_{i,P-1},0,\ldots,0]^{\mathrm{T}}\in\mathbb{C}^{\blck}$ denotes the time-domain channel-taps of the $i$-th active user and $\mathbf{F}\in\mathbb{C}^{N\times N}$ denotes a DFT matrix. The second expression in \eqref{eq:freq_multitap} shows that the multi-tap propagation condition leads to the reception of frequency-modulated copies of transmitted sequences. In Section~\ref{sec:dictionary_design} we discuss how the properties of the Gabor dictionaries can be leveraged to mitigate the interference caused by the above effect.}
\smallskip

\noindent \textit{\textbf{Discussion:}} The system model considered here can be put in the context of \cite{polyanskiy17}, by considering a setup where the system users share a total number of $\ntot$ channel resources that are split into $V$ (resource) blocks, each of size $\blck=\ntot/V$. In the U-RA setting with a common dictionary, an active user would then randomly select one of the $V$ resource blocks for transmission of the codeword of length $\blck$. At the receiver side, the decoding process is performed independently over each block, with the aim to retrieve the messages transmitted by the active users in the corresponding block. Similarly, in the S-RA setting where the users apply separate (user-specific) dictionaries, the total number of system users (say $K_{\mathrm{tot}}$), would be split into groups of $K$ users, where a group of $K$ users is "configured" to share one of the $V$ resource blocks of size $\blck$ channel uses. We remark that in both settings, a more general system model is conceivable according to which the active users spread their messages over $\rho\geq 1$ resource blocks, such that coding can also be applied over several resource blocks. This would, for example, be relevant to a block-fading scenario where time/frequency diversity may be exploited by coding over several fading blocks. We note, however, that this would require some additional level of coordination in the system, as the system users should be divided into groups such that the users from the same group are configured to share the same $\rho$ resource blocks.

\vspace{-7pt}
\section{Code Design} \label{sec:Code_Design}

Consider the general encoding model \eqref{eq:system_model} according to which the user maps the information message $W\in [2^{\info}]$ to the binary codeword $\mathbf{c}\in\mathcal{C}\subset\{0,1\}^{\mw}$, which is then mapped to the transmit vector 
$\mathbf{s}_i=\mathbf{A}_i\mathbf{c}_i$. In the following we discuss the design of the code $\mathcal{C}$ for both scenarios (common and separate dictionaries). Before we proceed, we will need some preliminaries.
\vspace{-7pt}
\subsection{Preliminaries}
\begin{defn} \textit{(Constant-weight code)}
    An $(n, m, w, d)$ binary constant-weight (CW) code is a set of $m$ binary $n$-tuples of Hamming weight $w$ such that the pairwise overlap (maximum number of coincident $1$'s for any pair of codewords) does not exceed $d$. Any $(n, m, w, d)$ binary CW code can be described by an $m\times w$ incidence matrix on $ \left\{1,\cdots, n\right\}$ such that for every $a\in \left\{1, \cdots, m\right\}$, the row $[s(a,1), \cdots, s(a,w)]$ gives the locations of the $w$ 1's in the $a$-th codeword. The set of all CW codes with parameters $n$, $m$ and $w$ and $d$ is denoted $\mathcal{CW}(n,m,w,d)$. 
\end{defn}

\begin{defn} \textit{(OR superposition)}
    Consider a set $\mathcal{A}=\{\mathbf{c}_1, \ldots, \mathbf{c}_p\}$ consisting of $p$ binary vectors of length $n$. We define the \textit{OR superposition} of these vectors as the binary vector
    \begin{equation}
        \mathbf{z}=f(\mathcal{A})\triangleq \mathbf{c}_1\vee \mathbf{c}_2\vee \cdots \vee \mathbf{c}_p.
    \end{equation}
Also, a binary vector $\mathbf{c}$ is said to be \textit{included} in a binary vector $\mathbf{d}$ if and only if $\mathbf{c}\vee \mathbf{d}=\mathbf{d}$.  
\end{defn}

\begin{defn} \textit{(Uniquely decipherable code \cite{Kautz})} 
    The binary code $\mathcal{C}$ with codeword length $n$ and size $m$ is a \textit{uniquely decipherable code} of order $p$ if every OR superposition of up to $p$ different codewords is distinct from every other sum of $p$ or fewer codewords. The set of all uniquely decipherable codes with parameters $n$, $m$ and $p$ is denoted $\mathcal{U}(n,m,p)$.\label{def:uniquely_decipherable_code}
\end{defn}

\begin{defn} \textit{(Disjunctive code \cite{AbdulJabbar88})}
    The binary code $\mathcal{C}$ with codeword length $n$ and size $m$ is a \textit{disjunctive code} or also \textit{zero-false-drop} of order $p$ if each subset $\mathcal{A}\subseteq \mathcal{C}$ of size $\vert \mathcal{A}\vert \leq p$ has the property that $\forall \mathbf{c}\in \mathcal{A}$ we have  
        $\langle \mathbf{c}, f(\mathcal{A})\rangle =\vert\mathbf{c}\vert_{\mathrm{H}}$, 
    but for all other codewords $\tilde{\mathbf{c}}\in\mathcal{C}\setminus \mathcal{A}$ we have
        $\langle \tilde{\mathbf{c}}, f(\mathcal{A})\rangle \leq  \vert\tilde{\mathbf{c}}\vert_{\mathrm{H}}-1$. 
    In the above, for two binary vectors $\mathbf{x}, \mathbf{y}$ we have defined $\langle \mathbf{x}, \mathbf{y} \rangle$ to be the \textit{correlation}, i.e. the pairwise overlap (the number of positions where both $\mathbf{x}$ and $\mathbf{y}$ have $1$'s). The set of all disjunctive codes with parameters $n$, $m$ and $p$ is denoted $\mathcal{D}(n,m,p)$. \label{def:disjunctive_code}
\end{defn}
\begin{defn} \textit{(Protocol sequence)}
    The binary code $\mathcal{C}$ with length $n$ and size $m$ is a \textit{protocol sequence} of order $p$ if any set $\mathcal{A}\subset \mathcal{C}$ of size $p$ or less has the property that any $\mathbf{c}\in \mathcal{A}$ has at least one position where all other codewords in $\mathcal{A}$ have a zero. 
    The set of all protocol sequences with parameters $n$, $m$ and $p$ is denoted as $\mathcal{P}(n, m, p)$.   
\end{defn}
\vspace{-7pt}

\subsection{Shared Dictionary}\label{sec:encoding_common_dictionary}

Consider the U-RA setting from Section~\ref{sec:common_dictionary}, for which, we recall, the received signal reads
\begin{equation*}
    \mathbf{y}=\ACD\mathbf{x}+\noise,
\end{equation*}
with $\mathbf{x}=\sum_{i\in\Ia} h_i\mathbf{c}_i$. This channel can be viewed as a concatenation of an \textit{inner} channel $\mathbf{x}\rightarrow \ACD\mathbf{x} + \noise$, and an \textit{outer} multiple access channel, $(\mathbf{c}_{i_1}, \ldots,\mathbf{c}_{i_{\Ka}})\rightarrow \mathbf{x}$, where $(\forall k\in[\Ka])\; i_k\in\Ia$. Following \cite{Fengler2021}, we will refer to the corresponding encoder and decoder as inner and outer encoder/decoder. To code jointly for the concatenated channel, we employ a binary constant-weight (CW) code construction obtained by concatenating a code with a non-binary alphabet with a \textit{pulse position modulation} (PPM) code. As discussed, the construction is motivated by considering the simplified inner channel, where instead of estimating the full signal $\mathbf{x}$, the inner decoder estimates its support,
\begin{equation}
    \mathrm{supp}(\mathbf{x})_j=\begin{cases}
        1, \vert x_j\vert\geq \theta,\\
        0,\: \mathrm{otherwise}
    \end{cases},
\end{equation}
with $\theta$ being an appropriately selected threshold. With this simplification, the outer channel becomes a (noisy) binary input OR-MAC channel, $\bigvee\limits_{i\in\mathcal{I}_{\mathrm{a}}} \mathbf{c}_{i}\rightarrow \mathrm{supp}(\mathbf{x})$. 

\noindent\textit{\textbf{CW Codes as Disjunctive Codes:}} According to Definition~\ref{def:uniquely_decipherable_code}, for a uniquely decipherable code of order $p$, every OR superposition of up to $p$ different codewords is distinct from every other sum of $p$ or fewer codewords. In \cite{Kautz} it has been shown that the class of disjunctive (zero-false-drop) codes (Definition~ \ref{def:disjunctive_code}) of order $p$ is a subset of the class of uniquely decipherable codes of the same order, $\mathcal{D}\left(n,m,p\right)\subseteq \mathcal{U}\left(n,m,p\right)$. 

We note that this relation suggests that the condition that a code is disjunctive (zero-false-drop) is more restrictive than the condition that the code is uniquely decipherable. In a follow-up on \cite{Kautz}, \cite{AbdulJabbar88} establishes a relation between CW codes with weight $w$ and correlation $d$, and disjunctive codes of order $\lceil\frac{w}{d}\rceil-1$ as    
\begin{equation}
    \mathcal{CW}\left(n,m,w,d\right)\subseteq \mathcal{D}\left(n,m,\left\lceil \frac{w}{d}\right\rceil -1 \right),
    \label{eq:CW_Disjunctive_Relation}
\end{equation}
where $\lceil x \rceil$ denotes the smallest integer greater than or equal to $x$. The relation (\ref{eq:CW_Disjunctive_Relation}) is established via the set of protocol sequences of order $p$, for which we have (see \cite{AbdulJabbar88})
\begin{align}
    \mathcal{D}(n,m,p-1)&=\mathcal{P}(n,m,p)\\
    \mathcal{CW}\left(n,m,w,d\right)&\subseteq \mathcal{P}\left(n,m,\left\lceil \frac{w}{d}\right\rceil \right).
\end{align}
The relation (\ref{eq:CW_Disjunctive_Relation}) suggests that, with the right parameterization,  CW codes yield disjunctive codes of the required order that guarantee separation of the users' codewords from their OR-superposition. While different families of CW codes can be used for this purpose, with examples including codes based on finite geometries, we address a construction obtained by concatenating a pulse-position-modulation (PPM) code and a Reed-Solomon (RS) code, which is appropriate for the task of finding codes with a short length $n$ for a fixed order $p$ and size $m$ \cite{AbdulJabbar88}.\smallskip 

\noindent\textit{\textbf{CW Code Construction:}}
Let $(q, q, 1, 0)$ be a $\mathrm{PPM}(q)$ code, i.e. a binary constant-weight code that consists of all binary $q$-vectors of unit weight. If we concatenate the $\mathrm{PPM}(q)$ code with an outer code with alphabet size $q$, blocklength $n'$, size $m'$ and minimum distance $d'$, we will obtain an $(n, m, n,  d)$ CW code with blocklength $n=n'q$, size $m=m'$, weight $w=n'$, and correlation $d=n'-d'$. In the following we address a construction where a CW code is obtained by concatenating a PPM code with an RS code. For a prime power $q$, and $k'<n'\leq q$, the corresponding RS code $\mathrm{RS}(n',k')$ has $k'^q$ codewords with minimal distance $d'=n'-k'+1$, as RS codes are maximum distance separable (they meet the Singleton bound).

From \eqref{eq:CW_Disjunctive_Relation} we have that the resulting CW code is a disjunctive code of order 
\begin{equation*}
p=\left\lceil \frac{w}{d} \right\rceil-1=\left\lceil \frac{n'}{k'-1} \right\rceil-1,    
\end{equation*}
meaning that any combination of $p$ (or less) codewords can be resolved from their OR-superposition. In other words, as long as the number of active users overlapping on the shared resources does not exceed $\lceil \frac{n'}{k'-1} \rceil-1$, their corresponding messages can be resolved without ambiguity. We will see that, in practice, this is rather the "worst case" scenario, i.e. the number of overlapping users that can be resolved (under the overall target error probability) is typically greater.   

\subsubsection{An example}
Consider the CW code obtained by the concatenation of a $\mathrm{PPM}(8)$ code and a systematic Reed-Solomom $RS(6,2)$ code over $\mathbb{GF}(8)$ with a generator matrix 
\begin{equation*}
    G_{\mathrm{RS}}=\begin{small}
    \left(\begin{array}{c} 1\\0\end{array} \begin{array}{c} 0\\1\end{array} \begin{array}{c} 6\\4\end{array} \begin{array}{c} 1\\1\end{array} \begin{array}{c} 6\\5\end{array} \begin{array}{c} 7\\5\end{array}\right)
    \end{small}.
\end{equation*}
Further, let the mapping from the $8$-ary symbols to the binary $\mathrm{PPM}$ codewords be given by $0\rightarrow (1\:0\: 0\: 0\: 0\: 0\: 0\: 0)^{\mathrm{T}}$, $1\rightarrow (0\:1\: 0\: 0\: 0\: 0\: 0\: 0)^{\mathrm{T}}$, $\ldots$ $7\rightarrow (0\:0\: 0\: 0\: 0\: 0\: 0\: 1)^{\mathrm{T}}$. 
The resulting binary CW code $\mathcal{C}_{\mathrm{CW}}=\mathrm{PPM}(2^3)\circ \mathrm{RS}(6,2)$ is of dimension $n=48$ and consists of $m=64$ codewords of weight $w=6$ and correlation $d=1$. According to \eqref{eq:CW_Disjunctive_Relation}, it is also a disjunctive code of order $p=\left\lceil \frac{6}{1}\right\rceil - 1=5$. As a consequence, any combination of codewords (given that they are different) of up to $p=5$ users can be resolved from their OR superposition. Take for example the scenario with three active users transmitting three different RS codewords, $   \mathbf{c}_1^{\mathrm{RS}}=(5\:0\:3\:5\:3\:6)$, $\mathbf{c}_2^{\mathrm{RS}}=(4\:2\:6\:6\:4\:0)$, and $\mathbf{c}_3^{\mathrm{RS}}=(1\:0\:6\:1\:6\:7)$   
respectively, giving rise to the codewords $\mathbf{c}_1,\mathbf{c}_2,\mathbf{c}_3\in\mathcal{C}_{\mathrm{CW}}(48, 64, 6, 1)$. Now, let us arrange the resulting OR superposition $\mathbf{c}^{\mathrm{OR}}=\mathbf{c}_1\vee\mathbf{c}_2\vee\mathbf{c}_3$ in a $8\times 6$ matrix, following the notation in \cite{AbdulJabbar88}\smallskip  
\begin{equation*} 
    \mathbf{C}^{\mathrm{OR}}=\begin{small}\left(
    \begin{array}{c} 0\\1\\0\\0\\0\\1 \end{array} 
    \begin{array}{c} 1\\0\\0\\1\\0\\0 \end{array} 
    \begin{array}{c} 0\\1\\0\\0\\0\\0 \end{array} 
    \begin{array}{c} 0\\0\\1\\0\\1\\0 \end{array} 
    \begin{array}{c} 1\\0\\0\\0\\1\\0 \end{array} 
    \begin{array}{c} 1\\0\\0\\1\\0\\0 \end{array}
    \begin{array}{c} 0\\0\\1\\1\\1\\1 \end{array}
    \begin{array}{c} 0\\0\\0\\0\\0\\1 \end{array}
    \right)^{\mathrm{T}}\end{small}.
\end{equation*}

\begin{figure}
\centering
\includegraphics[width=.67\linewidth]{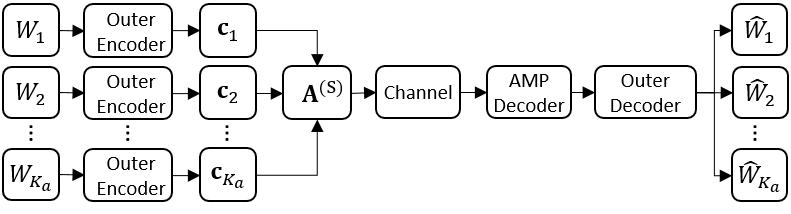}
\caption{\textcolor{black}{Flow chart of U-RA coding/decoding procedure.}}
\label{fig:fcunsourced}
\end{figure}

Due to the systematic form of the $\mathrm{RS}(6,2)$ code over $\mathbb{GF}(8)$, we observe from the first two columns of $\mathbf{C}^{\mathrm{OR}}$ that the sets of $8$-ary symbols $\mathcal{S}_1=\{1,4,5\}$ and  $\mathcal{S}_2=\{0,2\}$ are contained in the OR superposition. From the Cartesian product $\mathcal{S}_1\times \mathcal{S}_2=\{(1, 0), (1,2), (4,0), (4,2), (5,0), (5,2)\}$, we can then correlate $\mathbf{C}^{\mathrm{OR}}$ with the $6$ codewords from the CW code that have the pairs of symbols from $\mathcal{S}_1\times \mathcal{S}_2$ at the first two positions in the $\mathrm{RS}(6,2)$ code. Based on Definition 4 (disjunctive codes), the result of the correlation for $\mathbf{c}_1$, $\mathbf{c}_2$ and $\mathbf{c}_3$ will be $6$ (i.e. the same as the weight of the CW code), while for the remaining codewords the correlation will be strictly less. Hence, in this way, the transmitted codewords can be resolved from the OR superposition without ambiguity. In Fig.~\ref{fig:fcunsourced} we summarize the communication steps of the U-RA setting in a flow-chart.      
\vspace{-7pt}
\subsection{User-Specific Dictionaries}
\begin{figure}
\centering
\includegraphics[width=.67\linewidth]{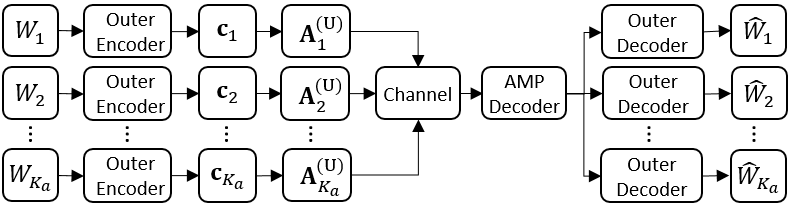}
\caption{\textcolor{black}{Flow chart of S-RA coding/decoding procedure.}}
\label{fig:fcsourced}
\end{figure}

Consider the S-RA setting from Section~ \ref{sec:separate_dictionaries}. When active, the users map their binary codewords (vectors) $\mathbf{c}\in \mathcal{C}\subset \{0,1\}^{\mw}$ on the considered resource block by employing separate (i.e. user-specific) dictionary matrices. We recall the resulting receive signal
\begin{equation*}
    \mathbf{y}=\ASD\mathbf{x}+\noise
    \label{eq:separate_dictionaries_1}
\end{equation*}
where, as defined in Section~\ref{sec:separate_dictionaries}, $\ASD=[\mathbf{A}_1,\ldots, \mathbf{A}_K]\in\mathbb{C}^{N\times \Ktot\mw}$ is the matrix that concatenates the dictionary matrices of the $K$ users, and $\mathbf{x}\in\mathbb{C}^{\Ktot\mw}$ is defined in \eqref{eq:separate_dictionaries}. Considering that $\Ka$ out of the $\Ktot$ users are active over the considered resource block, this channel can be viewed as a concatenation of an inner channel,  
    $\mathbf{x}\rightarrow \ASD\mathbf{x}+\noise$, 
and a bank of parallel outer channels, 
    $\mathbf{c}_i\rightarrow \mathbf{x}_i, \forall i\in\mathcal{I}$. 
To code over this channel, we propose to use essentially the same concept as in the U-RA  setting, i.e. to apply a binary CW code obtained by concatenating a code with a non-binary alphabet (such as an RS code), with a PPM code. The interpretation here is, however, different, as the role of the outer non-binary code is to correct the section errors at the output of the inner decoder, rather than to provide user separation. As result, also the parameterization of the CW code obtained by the above concatenation, is different in general. \textcolor{black}{In Fig.~\ref{fig:fcsourced} the communication steps of the S-RA setting are summarized in a flow-chart.}
\vspace{-7pt}
\subsection{Discussion and Related Work}
The proposed approach has certain similarities, but also some important differences to the current approaches from the literature. For example, in the S-RA setting with separate dictionaries, if we restrict the CW code construction to a single PPM code, without the concatenation with the outer algebraic code (Reed-Solomon code in our example), the transmission scheme resembles the one in \cite{Larsson_1} where each user selects one pilot sequence out of a set of $M=2^r$ sequences to convey $r$ bits of information. \textcolor{black}{We note, that the coding scheme from \cite{Larsson_1} leads to a detection problem with complexity that is exponential in $r$ which is effectively reduced by the CW code construction due to the structure of the outer algebraic code}. Besides the use of the outer algebraic code, when compared to \cite{Larsson_1}, which relies on the "conventional" AMP implementation, here we consider a more general inference procedure that accounts for the structure of the applied CW code to perform user activity detection, symbol detection and decoding. 

When applied to the U-RA setting with a common dictionary, our transmission scheme has analogy with the scheme proposed in \cite{ordentlich17}, which considers a concatenation of an inner binary linear code, and an outer code whose goal is to enable the receiver to recover the individual messages that participated in the modulo-2 sum. We note, however, that \cite{ordentlich17} considers a Gaussian channel model where, as result, the receiver can first decode the modulo-2 sum of all codewords within the same block (compute-and-forward (CoF) phase), which produces a binary adder channel (BAC) at the output of the modulo-sum decoder. The applied outer code should then enable the receiver to recover the individual messages that participated in the modulo-2 sum (BAC phase). As we are also considering a fading channel model and non-coherent communication without channel estimation, the users can not perform precoding that is necessary for the CoF decoder to output the modulo-2 sum. In addition, in our non-coherent approach, after the AMP module the receiver has effectively access to the output of a BAC with addition over the reals, rather than a modulo-2 BAC. In this respect, our scheme is conceptually  similar to \cite{Fengler2021}, where the inner channel is simplified by restricting the AMP module to output the support, rather than estimating the full signal $\mathbf{x}$. With this, the outer channel becomes a binary input OR-MAC, and the applied CW code in our scheme enables the receiver to separate the individual users, which is similar to the way that the outer tree code is applied in~\cite{Amalladinne2020}. \textcolor{black}{Besides the introduction of CW codes in the massive RA context, our scheme also differs from~\cite{Fengler2021} and~\cite{Amalladinne2020} in the application of Gabor dictionaries, which allows for shorter codeword sizes, as discussed in Section~\ref{sec:results}. }

\vspace{-7pt}
\section{Decoding}\label{sec:decoding}

In the following we provide high-level description of the decoding procedure for both considered scenarios. Details on the used algorithms are provided in Appendix~\ref{sec:appendix}-\ref{sec:appendix_outer_decoder}.\smallskip

\noindent \textit{\textbf{Shared Dictionary:}} In the U-RA scenario with common dictionary, the decoder represents a concatenation of an inner and outer decoder. The inner decoder is realized using an instance of the \ac{AMP} algorithm \cite{kim2011belief} which treats the entries of the vector $\mathbf{x}$ in \eqref{eq:CD} as independent, and outputs soft information in the form of log-likelihood ratios for the entries of the support of $\mathbf{x}$. The details of the inner decoder are provided in Appendix~\ref{sec:appendix} and summarized in Algorithm~\ref{alg:inner_decoder}. 
The outer decoder takes this soft information as input and outputs a list of codewords from the employed code $\mathcal{C}$ (in our case a constant-weight code), corresponding to the users active over the respective resource block. The decoding is based on the procedure in \cite{AbdulJabbar88} for decoding of superimposed codes over multiaccess OR channels and relies on the fact that the applied CW code $\mathcal{C}$ is disjunctive. The details of the outer decoder are provided in Appendix~\ref{sec:appendix_outer_decoder}. \smallskip

\noindent \textit{\textbf{User-Specific  Dictionaries:}}
The code construction in the scenario with separate dictionaries admits an iterative Bayesian user activity detection and decoding procedure. As for the scenario with common dictionaries, the decoding procedure is composed of an inner and outer decoder between which we pass beliefs for the entries of the support of $\mathbf{x}$ until some convergence criterion is met. For the inner decoder we use the \ac{AMP} based decoder as described in Algorithm~\ref{alg:inner_decoder}. Before passing the beliefs to the outer decoder, we introduce two additional steps in which we leverage the separate dictionaries and information on the structure of the CW code to enhance user and symbol detection. We provide a detailed derivation of the additional steps in Appendix~\ref{sec:AD}. The outer decoder is realized by a \ac{SISO} decoder for the non-binary outer code (RS code in our case) from \cite{bimberg10} which is applied on each codeword individually for a fixed number of iterations or until some convergence criterion is met. The resulting beliefs are then passed as prior beliefs to the inner decoder in the subsequent iteration of the overall decoding procedure. The overall decoding procedure is summarized in Algorithm~\ref{alg:sourced_decoder}.\smallskip

\noindent\textcolor{black}{ \textit{\textbf{Decoding Complexity:}} The decoding complexity of the U-RA setting is given by the complexity of the \ac{AMP}-based inner decoder and the complexity of the CWC outer decoder and scales as $\mathcal{O}(M\log M + \Ka^k)$. For the S-RA setting, the decoding complexity is provided similarly and scales as $\mathcal{O}(K M\log KM + q\log q)$. The complexity of the U-RA scheme predominantly scales with $\Ka$ and the complexity of the S-RA setting with $K$.
} 

\vspace{-7pt}
\section{Dictionary Design}\label{sec:dictionary_design}

An important issue related both to the performance and the encoding/decoding complexity is the choice of the user dictionaries (i.e. sequence design). As the number of iterations is finite, the decoding complexity scales linearly with the size of the design matrix. With a Gaussian design matrix, the memory requirement is also proportional to the dimension as the entire matrix has to be stored, which could be a bottleneck in scaling the AMP decoder to work with large matrices. 

\noindent \textcolor{black}{\textbf{Remark:} We note that \ac{AMP} may have convergence issues for challenging problem dimensions, and non Gaussian sensing matrices \cite{nonBayesianAdFengler}. Although Gabor frames exhibit similar properties to Gaussian matrices \cite{deter}, in the following we will resort to damping to stabilize \ac{AMP} \cite{ampDamping}.}
\vspace{-15pt}

\subsection{Dictionary Design based on Gabor Frames}

To reduce the decoding complexity and the required memory, we consider a construction based on finite \textit{Gabor frames}.
Gabor frames arise naturally in many important application areas such as communications, radar, and signal/image processing and have been used in the context of random access in \cite{Calderbank}. Besides having excellent coherence properties, Gabor frames are attractive since (i) they are completely specified by a total of $\blck$ numbers that describe the seed vector, and can be effectively generated as time-frequency translates of the seed vector, and (ii) multiplications with Gabor frames can be efficiently carried out using algorithms such as the FFT.

Formally, a Gabor frame is the set of all time-frequency translates of a nonzero unit-norm seed vector $\bm{g}\in\mathbb{C}^{\blck}$.
Let $\mathbf{g}_k$
denote the $k$-circular shifted vector $\mathbf{g}$,
then the Gabor frame generated from $\bm{g}$ is an $\blck\times\blck^2$ block matrix of the form
\begin{equation}
    \boldsymbol{\Phi}=[\mathrm{diag}(\mathbf{g}_0)\mathbf{F}, \mathrm{diag}(\mathbf{g}_1)\mathbf{F}, \ldots, \mathrm{diag}(\mathbf{g}_{\blck-1})\mathbf{F}]. \label{eq:Gabor_frame}
\end{equation}
In practice, a Gabor frame based on the Alltop seed vector~\cite{Bajwa2010} is particularly attractive due to its coherence properties. Formally, for a prime $\blck\geq 5$, the Alltop seed vector is constructed as 
\begin{equation}
    \bm{g}=[1,\ldots,e ^{j2\pi \frac{(\blck-2)^3}{\blck}},e ^{j2\pi \frac{(\blck-1)^3}{\blck}}]/\sqrt{\blck}. \label{eq:Alltop_seed} 
\end{equation}
 The elements of a Gabor frame $\boldsymbol{\Phi}$ generated from the Alltop seed vector 
satisfy~\cite{Strohmer} 
 $   \mu (\boldsymbol{\Phi})\doteq \max_{\substack{i,j : i\neq j}}\vert \langle \boldsymbol{\phi}_i,\boldsymbol{\phi}_j\rangle\vert \leq 1/\sqrt{\blck}$. 
More precisely, this particular frame construction represents a union of $\blck$ orthonormal bases of $\mathbb{C}^\blck$, and the modulus of the inner products between frame elements takes on only the values $0$ and $1/\sqrt{\blck}$, $\vert\langle \boldsymbol{\phi}_{i},\boldsymbol{\phi}_{j}\rangle\vert\in\{0, 1/\sqrt{\blck}\}$, $\forall i\neq j$. Given \eqref{eq:Gabor_frame}, the dictionaries for the S-RA and U-RA scenario are constructed by selecting first $\Ktot M L$ and $ML$ sequences respectively, i.e., $\ACD=[\boldsymbol{\phi}_0,\ldots,\boldsymbol{\phi}_{\Ktot M}]$ and $\ASD=[\boldsymbol{\phi}_0,\ldots,\boldsymbol{\phi}_{M}]$.\smallskip

\textcolor{black}{We remark that sequences from Gabor frames exhibit interesting properties the can be leveraged to mitigate the effect of interference  encountered in frequency-selective fading channels. Indeed, with the Gabor frame construction in  \eqref{eq:freq_multitap}, the dictionaries for the S-RA and U-RA scenarios can be made robust against $P$-tap frequency-selective channels by  restricting the size of the transmit dictionary by selecting sequences from a Gabor-frame that are $P$ sequences apart. This is due to the fact that Gabor frames are block circulant as result of the construction via a time-frequency expansion of a seed sequence. Gabor frames can thus be leveraged to design dictionaries which are robust against frequency selective channels.} 
\vspace{-7pt}

\subsection{Discussion/Performance}
\begin{figure}[h!]
    \centering
    \includegraphics[width=.5\textwidth]{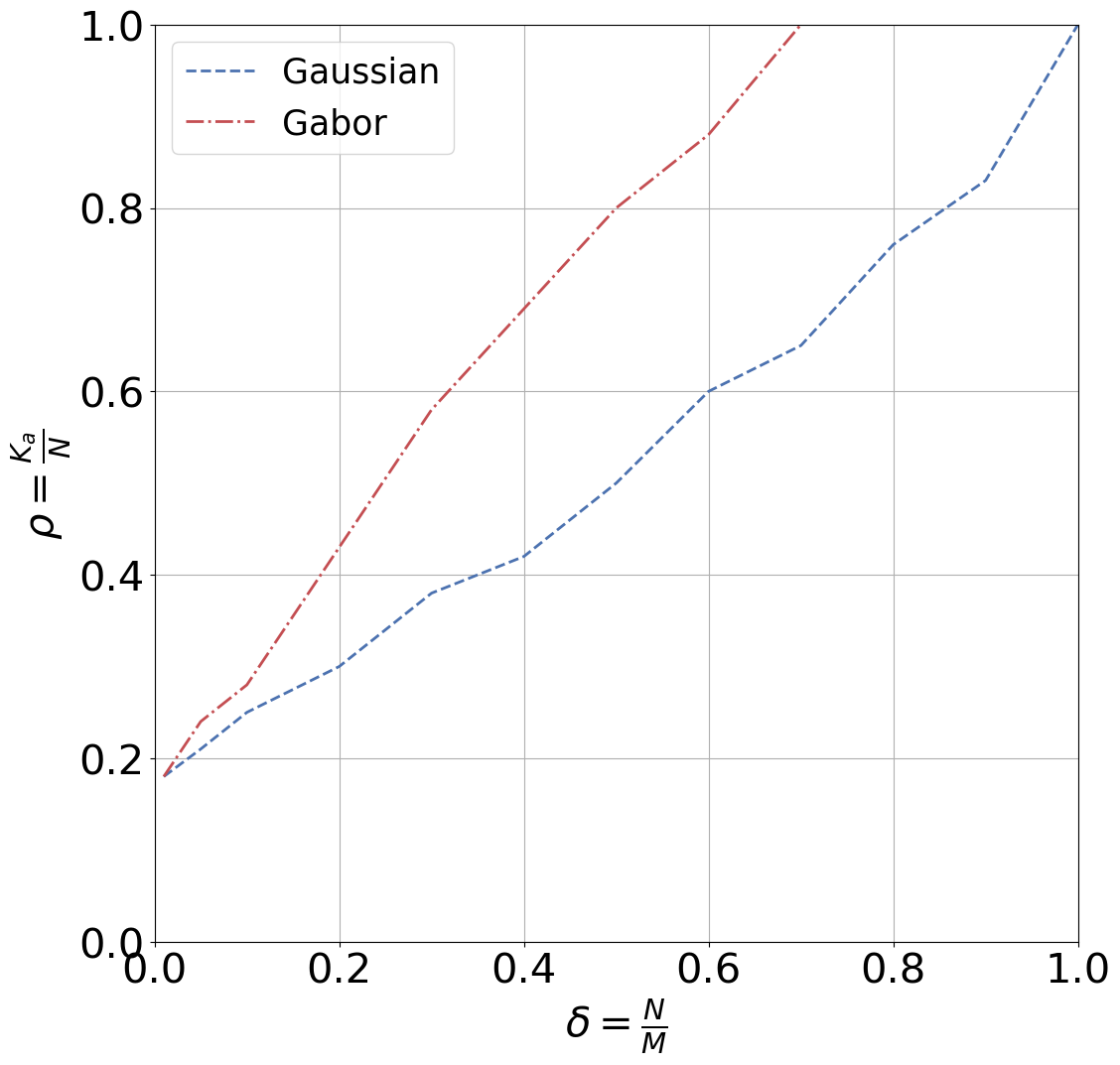}
    \caption{Empirical phase transitions of Gabor and Gaussian codebook with seq. length $N=257$.}
    \label{fig:phasetransition}
\end{figure}

The Gabor frame construction in \eqref{eq:Gabor_frame} holds interesting properties that make it amenable for the  dictionary design problem of interest here. Since the task of the inner decoder can be stated as an instance of compressive sensing (CS) reconstruction, the performance of Gabor frames can be analyzed by using tools from CS theory. 
It has been verified that Gabor frames with Alltop window have similar reconstruction performance as random Gaussian matrices \cite{deter}. 
In Figure~\ref{fig:phasetransition}, we depict the empirical phase transition for both Gabor and Gaussian codebook in the noiseless case, i.e., for the parameters below each phase transition line, the recovery of sparse vectors is perfectly achieved.
The phase transition suggests that for frame sizes in the order of $\approx \blck\times\blck^2$, the Gabor frame and Gaussian codebook behave in a similar fashion. However, we observe that \textit{truncated} Gabor frames, i.e., Gabor frames of size $\approx\blck\times M$ with $M<\blck^2$ tend to perform better than their Gaussian counterparts of same size, making them suitable candidates for our scenarios of interest.
\vspace{-7pt}

\vspace{-7pt}
\section{Numerical Simulations}\label{sec:results}

In the following we provide numerical performance evaluation of the proposed communication schemes from Section~\ref{sec:system_model}. Following the evaluation procedure from \cite{polyanskiy17}, we assume a total number of $\ntot$ channel resources that are split into $V$ (resource) blocks, each of size $\blck=\ntot/V$. In the U-RA setting with a common dictionary, an active user would then randomly select one of the $V$ resource blocks for transmission of the codeword of length $\blck$. At the receiver side, the decoding process is performed independently over each block, with the aim to retrieve the messages transmitted by the active users in the corresponding block. In the S-RA setting where the users apply separate (user-specific) dictionaries, the total number of system users $K_{\mathrm{tot}}$ is split into groups of $K$ users, where each group is "configured" to share one of the $V$ resource blocks of size $\blck$. 
Considering information messages of size $\info$ information bits that are encoded over $\blck$ channel uses, 
the performance is measured in terms of (i) the energy per-bit $\frac{E_b}{N_0}$ required to serve the active users with a fixed target error probability $P_e$, and (ii) the total number of active users $\bar{K}_{\mathrm{a}}$ that can be supported over the total number of channel resources $N_{\mathrm{tot}}$. 
\vspace{-7pt}
\subsection{Shared Dictionary (U-RA)}
Here we evaluate the performance of the proposed transmission scheme in the communication scenario from Section~\ref{sec:common_dictionary} where the users apply a common dictionary to map their binary codewords on the channel resources (U-RA setting). We consider both an AWGN and a Rayleigh fading scenario. In the AWGN scenario we assume single-antenna transmitters/receiver. In the fading scenario we assume single-antenna transmitters and multiple-antenna receiver. For the inner decoder we use Algorithm~\ref{alg:inner_decoder} with $T_{\mathrm{max}}=100$ iterations. The outer decoder is implemented as described in Appendix~\ref{sec:appendix_outer_decoder}. \smallskip

\subsubsection{AWGN channel}\label{sec:awgn_results}

\begin{figure}
	\centering
	\begin{subfigure}{.5\textwidth}
	  \centering
	  \includegraphics[width=.75\linewidth]{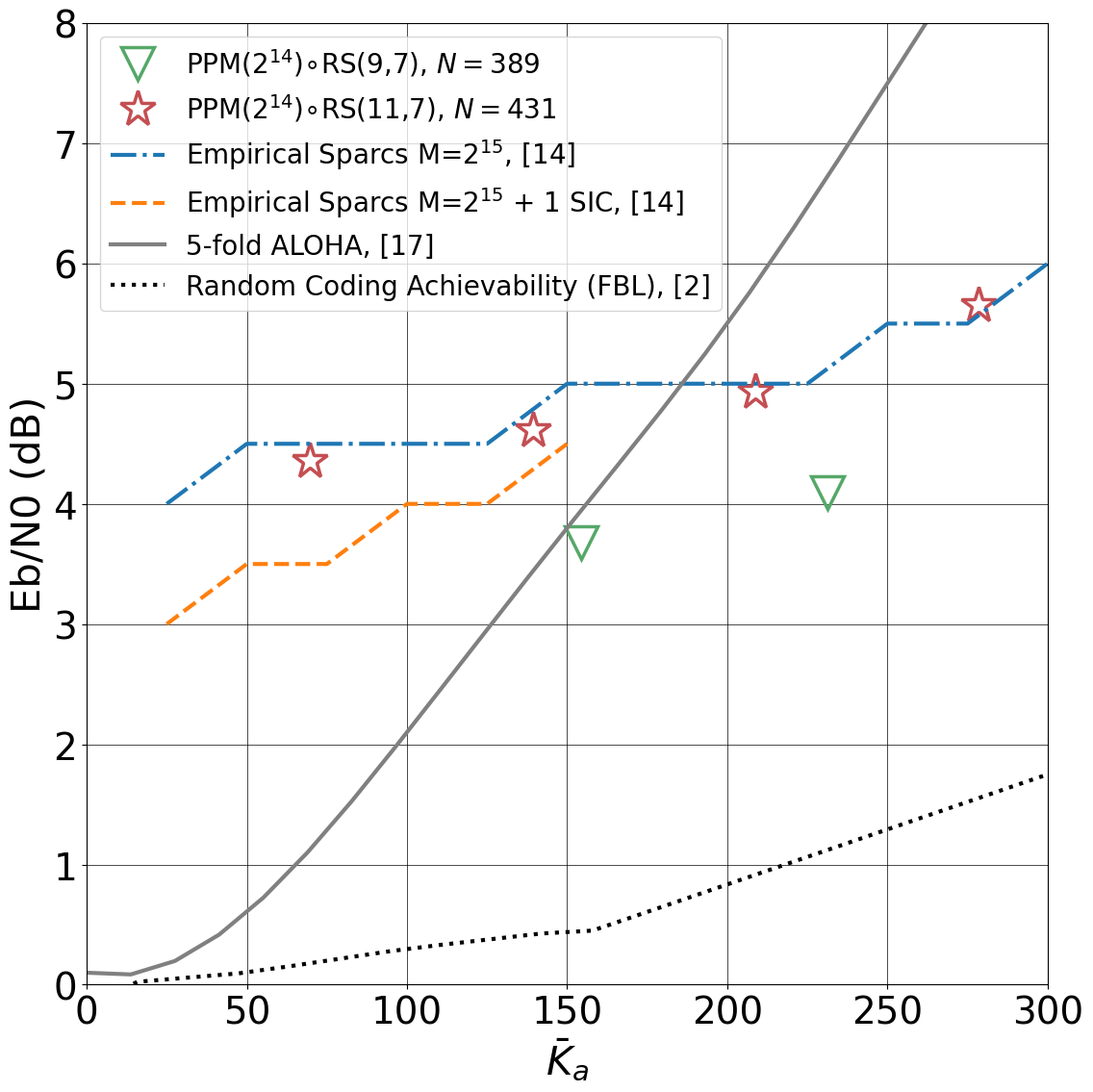}
	  \caption{}
	  \label{fig:unsourced_100_bits}
	\end{subfigure}%
	\begin{subfigure}{.5\textwidth}
	  \centering
	  \includegraphics[width=.75\linewidth]{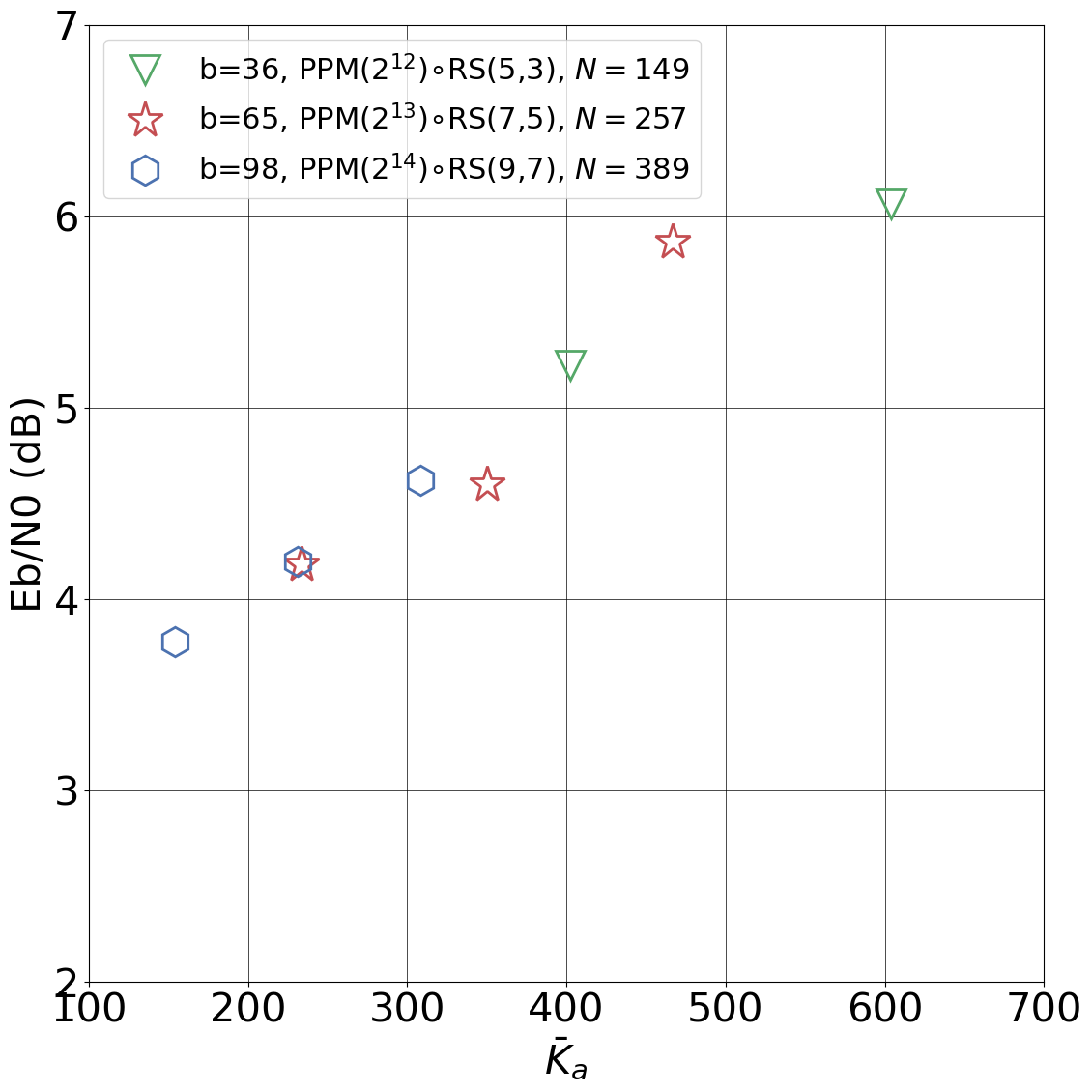}
	  \caption{}
	  \label{fig:unsourced_389_257_149}
	\end{subfigure}
	\caption{(a) Performance comparison of the CW code construction from Section~\ref{sec:Code_Design} with a Gabor dictionary, with state-of-the art codes from \cite{Fengler2021} and \cite{ordentlich17} (AWGN channel). The performance is expressed as the required $E_b/N_{0}$ (in dB) per active user vs. the total number of active users $\bar{K}_{\mathrm{a}}$ in the system, given a target error probability $P_e=0.05$. (b) Required $E_b/N_{0}$ per vs. the total number of active users $\bar{K}_{\mathrm{a}}$ for different message sizes (AWGN channel).}
\end{figure}

Fig.~\ref{fig:unsourced_100_bits} illustrates the performance of the coding scheme from Section~\ref{sec:encoding_common_dictionary}, where a user message of size $\info=98$ information bits is encoded over a block of $\blck$ channel uses. We test two designs for the CW code that prescribes which columns of the dictionary matrix are combined to produce the transmitted codeword. The first construction is obtained by concatenating a $\mathrm{PPM}(2^{14})$ code and a $\mathrm{RS}(11,7)$, resulting in a CW code with blocklength $n=11\cdot 2^{14}$.  The dictionary matrix $\ACD\in\mathbb{C}^{\blck\times\mw}$ is obtained by taking $M=n=11\cdot 2^{14}$ sequences from the Gabor frame in dimension $\blck=431$ based on the Alltop construction described in Section~\ref{sec:dictionary_design}. The second construction is obtained by concatenating a $\mathrm{PPM}(2^{14})$ code and a $\mathrm{RS}(9,7)$ code, resulting in a CW code with blocklength $9\cdot 2^{14}$. The dictionary matrix $\ACD\in \mathbb{C}^{\blck\times\mw}$ is obtained by taking $\mw=9\cdot 2^{14}$ sequences from the Gabor frame in dimension $\blck=389$ based on the Alltop construction. 
\vspace{-7pt}
We compare with the scheme based on the SPARCs construction from \cite{Fengler2021}, which applies the outer code from \cite{Amalladinne2020}, and against the schemes from \cite{ordentlich17}. For a fair comparison, we use the model from \cite{ordentlich17} and set the total number of channel uses to $N_{\mathrm{tot}}=30000$. Note that in \cite{Fengler2021}, the SPARCs construction is applied over the total number of channel uses (i.e. the codeword length is $N_{\mathrm{tot}}$), while here $N_{\mathrm{tot}}$ is split into $V$ resource blocks of size $N=N_{\mathrm{tot}}/V$, and each transmit codeword spans one such resource block. We observe that, for a similar message size ($98$ bits vs $100$ bits in the example in \cite{Fengler2021}), the $E_b/{N_{0}}$ performance (for a reliability target $P_e=0.05$) of the coding approach proposed here is comparable, or improves over the approach in \cite{Fengler2021}. In contrast to \cite{Fengler2021}, this is achieved with a \textit{considerably smaller size of the transmitted codewords} ($\blck\in\{431,389\}$ vs $N_{\mathrm{tot}}=30000$), which brings advantages in terms of transmission latency and/or bandwidth occupancy.\smallskip

\noindent\textit{\textbf{The impact of the messages size:}} In the following we compare the results from different parameterizations of the CW codes and sequence lengths $\blck$:
\begin{itemize}
\item \textit{Message size $\info=36$ bits:} The message is encoded over a single block of $\blck=149$ channel uses. The CW code is obtained by concatenation of a $\mathrm{PPM}(2^{12})$ code and an $\mathrm{RS}(5,3)$ code. The dictionary matrix $\ACD\in \mathbb{C}^{\blck\times\mw}$ is obtained by taking $\mw=5\cdot 2^{12}$ sequences from the Gabor frame of dimension $\blck=149$  based on the Alltop construction.

\item \textit{Message size $\info=65$ bits:} The message is encoded over a single block of $\blck=257$ channel uses. The CW code is obtained by concatenation of a $\mathrm{PPM}(2^{13})$ code and an $\mathrm{RS}(7,5)$ code. The dictionary matrix $\ACD\in\mathbb{C}^{\blck\times\mw}$ is obtained by taking $\mw=7\cdot 2^{13}$ sequences from the Gabor frame of dimension $\blck=257$ based on the Alltop construction.

\item \textit{Message size $\info=98$ bits:} The message is encoded over a single block of $\blck=389$ channel uses. The CW code is obtained by concatenation of a $\mathrm{PPM}(2^{14})$ code and an $\mathrm{RS}(9,7)$ code. The dictionary matrix $\ACD\in \mathbb{C}^{\blck\times\mw}$ is obtained by taking $\mw=9\cdot 2^{14}$ sequences from the Gabor frame of dimension $\blck=389$ based on the Alltop construction.
\end{itemize}
As depicted in Fig.~\ref{fig:unsourced_389_257_149}, with the increase of the blocklength $\blck$, and thus with the message size that can be supported, the $E_b/{N_{0}}$ required to meet the reliability target $P_e$ decreases. \textit{For shorter message sizes} (i.e. shorter sequence lengths), on the other hand, \textit{more active users can be served in the system at the cost of an (approx. linear) increase of the required $E_b/{N_0}$ (in dB) per user}.\smallskip 
 
\subsubsection{Block Rayleigh Fading Channel}
In the following we evaluate the performance of the proposed scheme with a common dictionary in a Rayleigh block-fading channel with coherence length no smaller than the block size $\blck$. We consider a non-coherent scenario, i.e. the fading realization in the block is unknown to both the transmitter and the receiver. We assume that the receiver has knowledge of the fading statistics. 
Furthermore, we assume single-antenna transmit devices and a multi-antenna receiver with $T$ antennas. 

\begin{figure}
	\centering
	\begin{subfigure}{.5\textwidth}
	  \centering
	  \includegraphics[width=.8\linewidth]{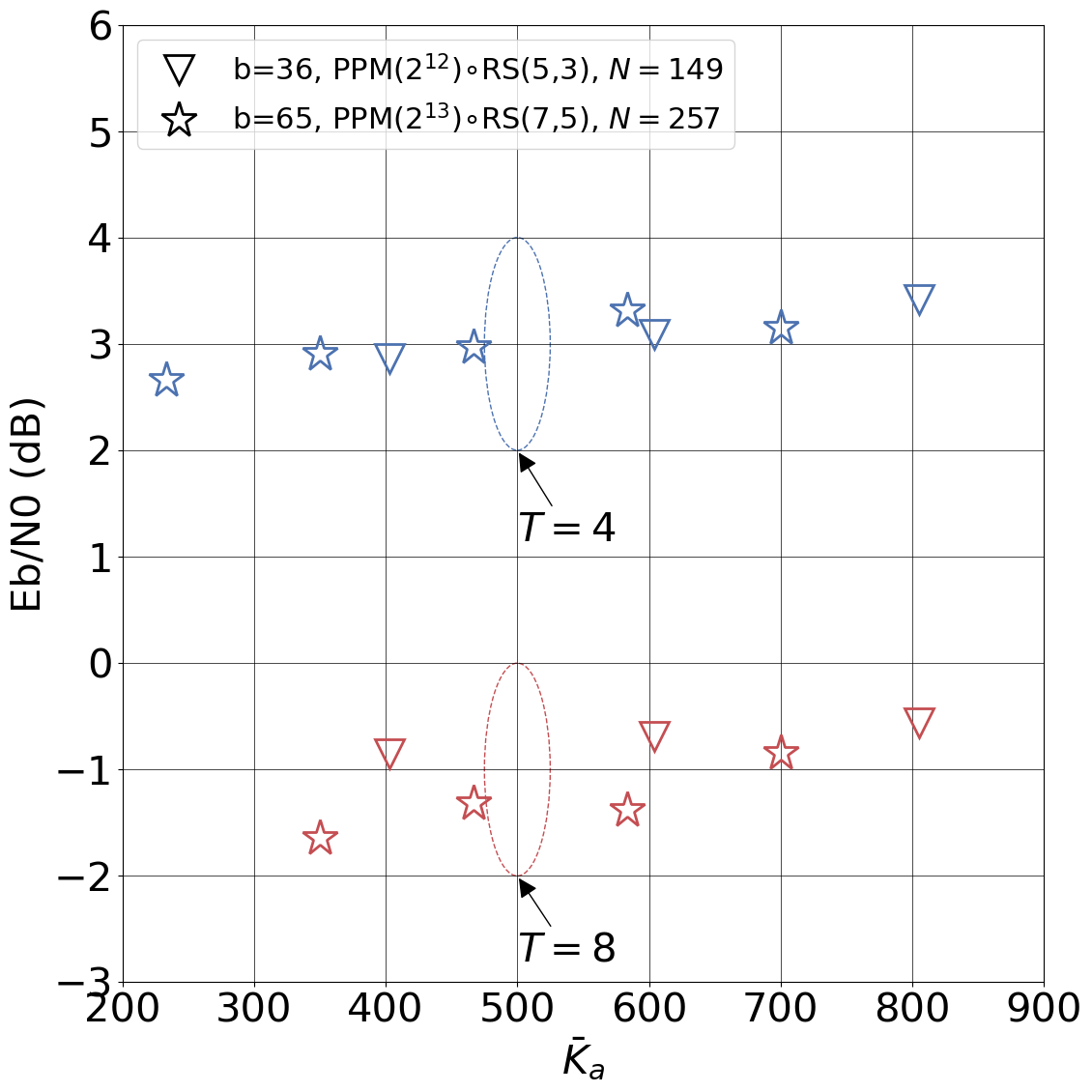}
	  \caption{Required $E_b/N_{0}$ vs $\bar{K}_{\mathrm{a}}$ (Rayleigh fading).}
	  \label{fig:unsourced_fading_1}
	\end{subfigure}%
	\hfill
	\begin{subfigure}{.5\textwidth}
	  \centering
	  \includegraphics[width=.8\linewidth]{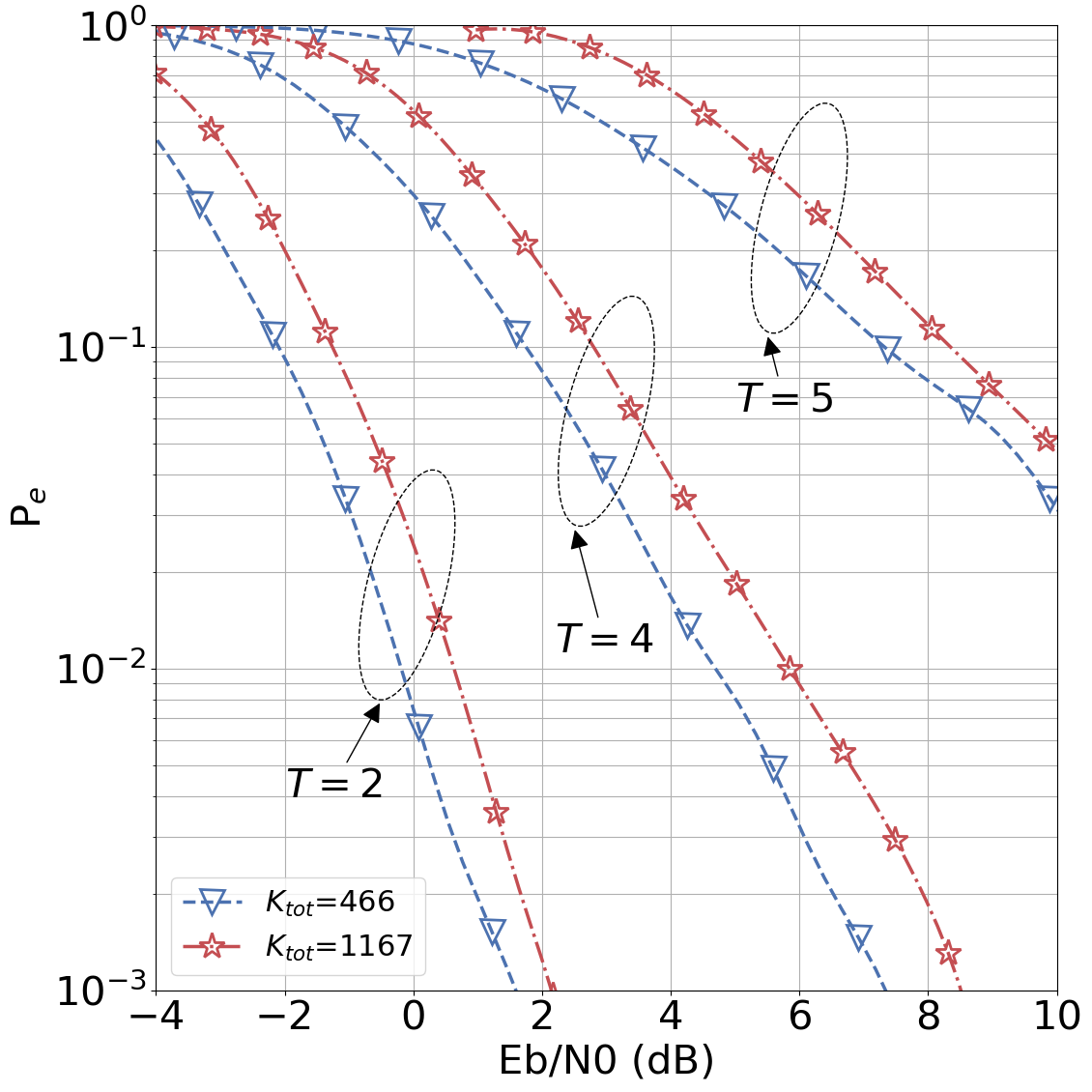}
	  \caption{Probability of error $P_e$ vs. $E_b/N_{0}$ (Rayleigh fading).}
	  \label{fig:multiantenna_257}
	\end{subfigure}
	\caption{Performance of the proposed scheme in a Rayleigh fading scenario with multiple-antenna receiver.}
\end{figure}

In Fig.~\ref{fig:unsourced_fading_1} we plot the required $E_b/{N_{0}}$ versus the number of active users $\bar{K}_{\mathrm{a}}$  in the system, for message sizes $\info=36$ and $\info=65$ bits, and corresponding code length $N=149$ and $N=257$ respectively. As before, we set $N_{\mathrm{tot}}=30000$ and $P_e=0.05$. 
We observe that, for both message sizes/code parameterizations, the number of active users that can be supported in the system increases with multiple receive antennas. Furthermore, the increase of the number of receive antennas from $T=4$ to $T=8$, results in a decrease of the required $E_b/N_{0}$ of approx. $4$ dB.

In Fig.~\ref{fig:multiantenna_257} we depict the per-user probability of error $P_e$ in a block-fading scenario as function of the required  $E_b/N_{0}$ to transmit a message of  size $\info=65$ bits in a system with $N_{\mathrm{tot}}=30000$ channel resources and two different system loads (number of active users), $\bar{K}_{\mathrm{a}}=466$ and $\bar{K}_{\mathrm{a}}=1167$. The active users encode their messages over blocks of length $\blck=257$, by using a CW code obtained from a concatenation of a $\mathrm{PPM}(2^{13})$ and a $\mathrm{RS}(7,5)$ code with a Gabor dictionary from the Alltop construction. The number of receive antennas $T$ is a parameter. 
\vspace{-7pt}
\subsection{User-Specific Dictionaries (S-RA)}

In the following we evaluate the S-RA communication scenario as described in Section~\ref{sec:separate_dictionaries}.
We consider an AWGN scenario with single-antenna transmitters/receiver, as well as a Rayleigh fading scenario with single-antenna transmitters and a multi-antenna receiver. 

\subsubsection{AWGN Channel}

\begin{figure}[!t]
	\centering
	\begin{subfigure}{.5\textwidth}
	  \centering
	  \includegraphics[width=.8\linewidth]{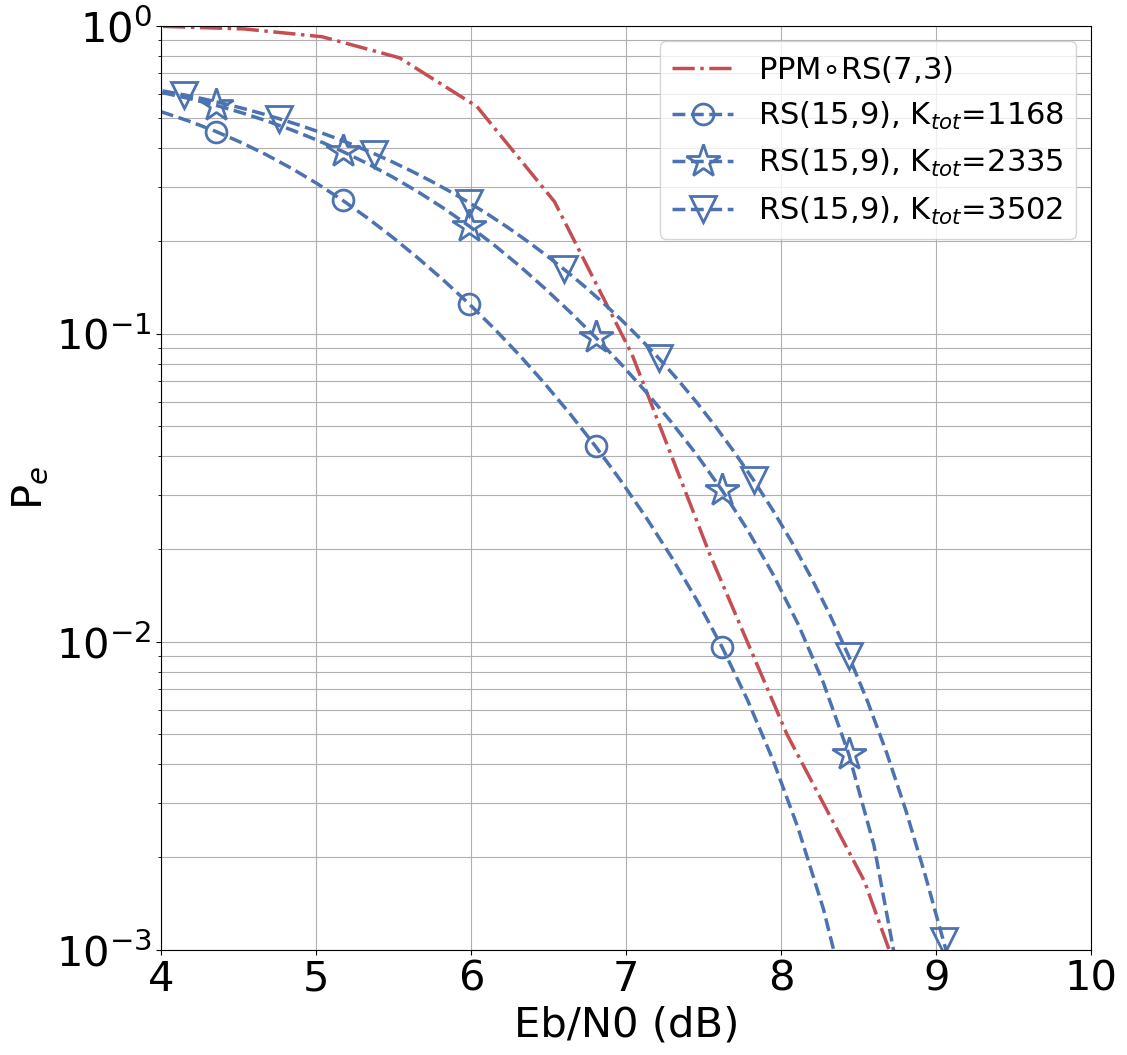}
	  \caption{AWGN channel.}
	  \label{fig:unsourced_vs_sourced_36bits}
	\end{subfigure}%
	\begin{subfigure}{.5\textwidth}
	  \centering
	  \includegraphics[width=.8\linewidth]{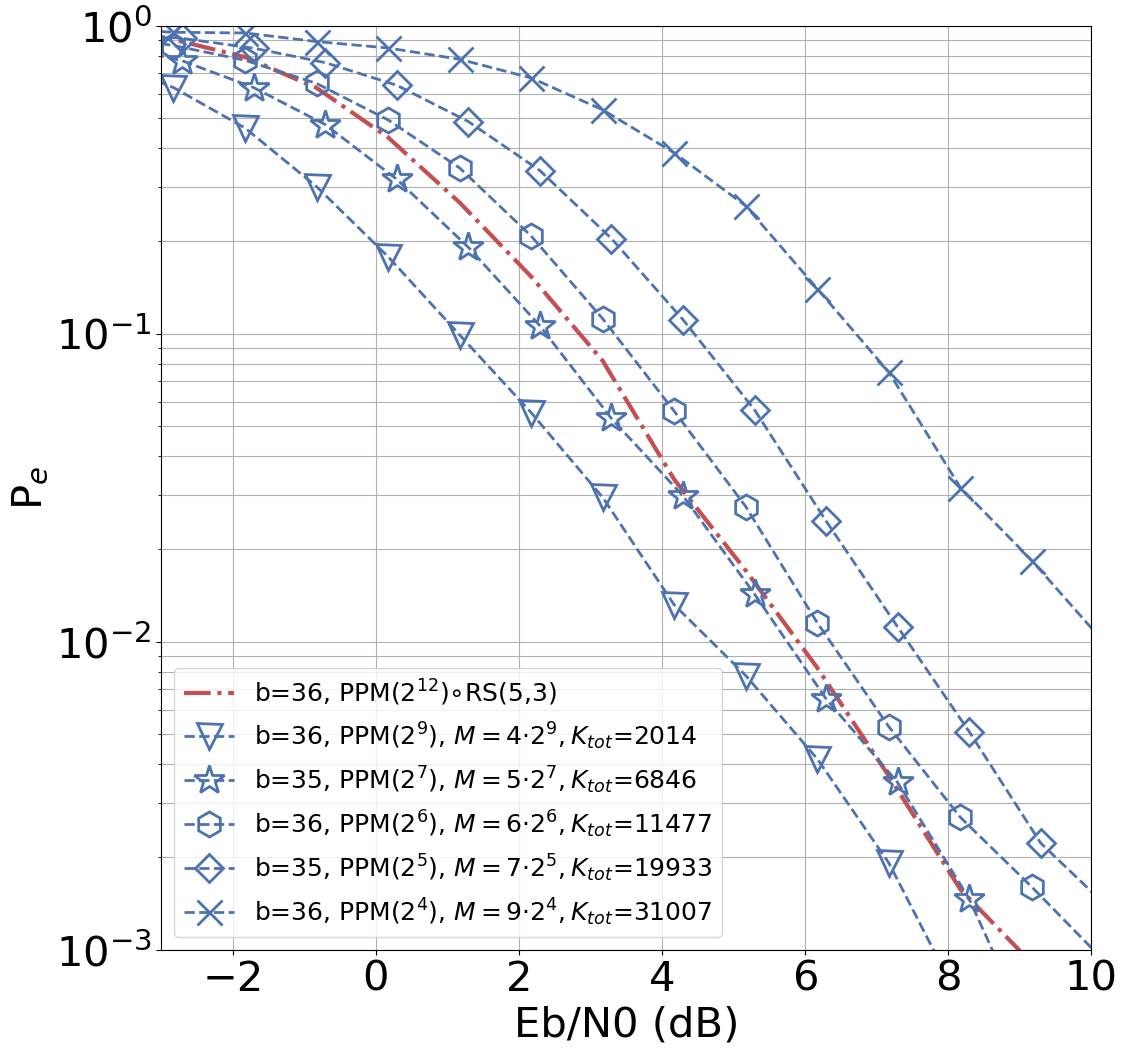}
	  \caption{Rayleigh fading channel with $T=4$ receive antennas.}
	  \label{fig:unsourced_vs_sourced_36bits_2}
	\end{subfigure}
	\caption{Performance of U-RA and S-RA in terms of error probability $P_e$ as function of  $E_b/N_{0}$.}
\end{figure}

In Fig.~\ref{fig:unsourced_vs_sourced_36bits} we aim to provide an assessment of the relative performance of the approach based on a common dictionary (U-RA) and the approach based on separated dictionaries (S-RA). In the simulations, we fix the message size to be $\info=36$ bits in both scenarios. As before, we assume a system with a total number of channel resources $N_{\mathrm{tot}}=30000$. The user messages are encoded over a block of $\blck=257$ channel uses. We note, however, that the relative assessment of the two approaches is qualitative rather than quantitative due to the conceptual differences between the two approaches: (i) in the S-RA setting, the overall number of system users ($K_{\mathrm{tot}}$) is fixed (and finite) and the use of separate dictionaries provides means for user identification; (ii) in the U-RA setting, on the other hand, there is no association between the transmitted messages and the user identities. As result, the overall number of system users ($K_{\mathrm{tot}}$) can be left out of the model, i.e. can be set to infinity.     

For the purpose of the comparison, in the U-RA setting with a common dictionary we use a CW code obtained by a concatenation of a $\mathrm{PPM}(2^{12})$ and an $\mathrm{RS}(7,3)$ code. The dictionary matrix $\ACD\in\mathbb{C}^{\blck\times \mw}$, which is shared among the users, is obtained by taking $M=7\cdot 2^{12}$ sequences from the $\blck=257$-dimensional Gabor frame based on the Alltop construction. In the S-RA setting we use the same Gabor frame, with the difference that the $257\times 257^{2}$ Gabor matrix is divided into $257$ sections of size $257$ sequences each, as given by  \eqref{eq:Gabor_frame} and \eqref{eq:Alltop_seed}. When $K\leq 257$ users are configured on the resource block of size $257$ channel uses, the $k$-th user ($1\leq k \leq K)$ is assigned the $k$-th section of the Gabor matrix. We take the first $15\cdot 16=240$ sequences from the section to obtain the dictionary matrix $\ASD_k\in \mathbb{C}^{257\times 240}$ that is used to map the user's message on the resource block of size $\blck=257$ resource elements. In the S-RA setting we evaluate the performance as function of the number of users $K$ configured on the same resource block of size $\blck=257$, where $K\in\{10, 20, 30\}$. The total number of system users is then $K_{\mathrm{tot}}=K\ntot/\blck$. In both scenarios we assume $K_{\mathrm{a}}=3$ active users within the resource block, which gives in total $\bar{K}_{\mathrm{a}}\approx350$ active users over the $\ntot=30000$ resources.

We observe that when the number of users $K$ configured on the same resource block of size $\blck=257$ is below $20$, corresponding to an overall number of system users $K_{\mathrm{tot}}\approx 2335$, it might be preferable to employ separate dictionaries (in terms of the required  $E_b/{N_{0}}$ to meet a target error probability $P_e=0.05$). In addition, we observe that, for higher $E_b/{N_{0}}$, using a common dictionary is associated with a steeper decrease of the error probability.

\subsubsection{Rayleigh Fading Channel}

Similarly to the AWGN case, in Fig.~\ref{fig:unsourced_vs_sourced_36bits_2} we provide a qualitative assessment of the relative performance of the two presented approaches (U-RA and S-RA) in a Rayleigh fading scenario. We fix the message size to $\info=36$ bits for the U-RA setting, and $\info=36$ or $\info=35$ bits in the S-RA setting, depending on the code parameterization. The user messages are encoded over a single block of $\blck=149$ channel uses. In both scenarios we consider a multiantenna receiver with $T=4$ receive antennas and $\Ka=6$ active users over the resource block, i.e. $\bar{\Ka}\approx 1208$ active users in total over the $\ntot=30000$ resources. 

In the U-RA setting we use a CW code obtained by a concatenation of a $\mathrm{PPM}(2^{12})$ and an $\mathrm{RS}(5,3)$ Reed-Solomon code. The common dictionary matrix $\ACD$ is obtained by taking $5\cdot 2^{12}$ sequences from the $\blck=149$-dimensional Gabor frame based on the Alltop construction. For the sourced access scenario we use the same  Gabor frame, with the difference that the Gabor matrix is divided into $K$ sections of size $\mw$ sequences each (by discarding some of the $149^2$ sequences to fit the  dimension $K\mw$, when necessary), where $K$ is the number of users configured on the resource block of size $N$ channel uses. The $k$-th user ($1\leq k \leq K)$ is assigned the $k$-the section of the Gabor matrix that serves as a unique dictionary $\ASD_k\in\mathbb{C}^{\blck\times \mw}$. 

In the S-RA setting we evaluate the performance for different numbers of system users $K$ configured on the same resource block of size $\blck=149$ channel uses. In the following, we consider only an inner $\mathrm{PPM}$ code (without and RS outer code), with the following parameterization:
\begin{itemize}
    \item $\mathrm{PPM}(2^9)$, $M=4\cdot 2^9$ (message size $\info=36$ bits), $K=10$ (corresponding to $K_{\mathrm{tot}}\approx 2014$).
    \item $\mathrm{PPM}(2^7)$, $M=5\cdot 2^7$ (message size $\info=35$ bits), $K=34$ (corresponding to $K_{\mathrm{tot}}\approx 6846$).
    \item $\mathrm{PPM}(2^6)$, $M=6\cdot 2^6$ (message size $\info=36$ bits), $K=57$ (corresponding to $K_{\mathrm{tot}}\approx 11477$).
    \item $\mathrm{PPM}(2^5)$, $M=7\cdot 2^5$ (message size $\info=35$ bits), $K=99$ (corresponding to $K_{\mathrm{tot}}\approx 19933$).
    \item $\mathrm{PPM}(2^4)$, $M=9\cdot 2^4$ (message size $\info=36$ bits), $K=154$ (corresponding to $K_{\mathrm{tot}}\approx 31007$).
\end{itemize}


We observe that when the number of users $K$ configured on the same resource block of size $\blck=149$ is below $K=34$, corresponding to an overall number of system users $K_{\mathrm{tot}}\approx 6846$, S-RA with user-specific dictionaries might be preferable over U-RA with a common dictionary (in terms of the required  $E_b/{N_{0}}$ to meet a target error probability $P_e=0.05$). The situation is reversed when increasing $K$ beyond this value. 
\vspace{-7pt}
\subsection{Summary of the results and discussion}
To summarize, in the above we presented simulation results for two settings: (i) unsourced random access (U-RA), where the users applied the same codebook, and (ii) "sourced" random access (S-RA), where the users applied separate codebooks that simultaneously identify the users. In both settings we assumed dictionary design based on a Gabor frame (Alltop construction), and codebook design based on a constant-weight (CW) code obtained by a concatenation of pulse-position modulation (PPM) and a Reed-Solomon (RS) code (with different parameterization depending on the setting). The figure of merit was the required $E_b/N_0$ (per user) to transmit a message of a fixed size, given a predetermined reliability target $P_e$. 

For the U-RA setting (AWGN scenario), the simulation results (Fig.~\ref{fig:unsourced_100_bits}) indicated that, compared to the SPARCs construction from from \cite{Fengler2021}, the $E_b/{N_{0}}$ performance 
of the coding approach proposed here is comparable, or improves over the approach in \cite{Fengler2021}. For a target message size  $\info\approx 100$ bits, this is achieved with a considerably smaller size of the transmitted codewords. We argue that this behavior is, in part, also a result of the statistical RIP properties of (truncated) Gabor frames, as indicated by the empirical phase transition depicted in Fig.~\ref{fig:phasetransition}. Further, to investigate the trade-off between the message size and the number of active users that can be supported by the system, in Fig.~\ref{fig:unsourced_389_257_149} we plotted the $E_b/N_0$ required to achieve a target reliability $P_e=0.05$ for  different (short) message sizes. We observed that the transmission of longer messages is more energy efficient, as the per-user $E_b/{N_{0}}$ required to meet the reliability target decreases with the message size. On the other hand, for shorter message sizes, we observed that more active users can be served in the system at the cost of an (approx. linear) increase of the required  $E_b/{N_0}$. Similar behavior was observed in the block fading scenario, where we have also quantified the influence of the number of receive antennas (Fig.~\ref{fig:unsourced_fading_1} and Fig.~\ref{fig:multiantenna_257}).

In the S-RA setting, the $E_b/N_0$ performance was evaluated as function of the number of system users configured on a resource block of certain size (Fig.~\ref{fig:unsourced_vs_sourced_36bits}). In general, it was observed that as the number of users configured on the same resource block increases beyond a certain threshold value, U-RA becomes more energy efficient than S-RA. The observation was replicated in the block Rayleigh fading scenario with multiple receive antennas (Fig.~\ref{fig:unsourced_vs_sourced_36bits_2}).

\textcolor{black}{To summarize, based on the presented results, we observe that the use of Gabor dictionaries in combination with constant-weight codes yields a form of coded modulation that seems to be a good match for the problem of non-coherent multiple access scenario with short messages. As result of the proposed combination, the transmission scheme can operate with comparably short transmit codeword sizes. This yields lower latency and smaller bandwidth occupancy, as well as lower complexity in block fading scenarios as messages can be transmitted over a single (or few) fading blocks.} 

\vspace{-7pt}
\section{Conclusion}\label{sec:conclusion}

In this paper we addressed a general framework for massive random access based on sparse superposition coding. According to the transmission scheme, 
the users convey information by linearly combining sequences from a predefined dictionary, with the addition that the sequence selection mechanism is prescribed by an appropriate error-correction code. 
For the receiver processing, we relied on an adaptation of the AMP algorithm that simultaneously accounts for the dictionary structure, as well as for the dependencies imposed by the involved error-correction code. The framework can be applied to the unsourced random access setting where all system users apply a common dictionary, as well as to the sourced random access setting where the users are assigned separate (i.e. user-specific) dictionaries. 

In the context of the joint transceiver design, a key observation was that the decoding process decomposes the communication channel into an inner channel, induced by the over-the-air linear mixture of the signals transmitted by the individual users, and an outer channel that is in effect a noisy binary input OR-MAC channel. To code for the inner channel we advocated a dictionary design based on Gabor frames, which, besides having excellent coherence properties, also yield significant reduction in both encoding/decoding complexity and storage requirements. To code for the outer channel we proposed the use of constant-weight codes, which, with the right parameterization, guarantee that the individual users can be resolved from their OR-superposition  with high probability. 
We performed numerical simulations for both the AWGN scenario and the non-coherent block-fading scenario without CSI at the transmitter/receiver side. The numerical simulations illustrated the potential of the approach to provide state-of-the art performance in terms of the required energy-per-bit (per user) to achieve a predefined reliability target, as well as in terms of the number of active users that can be simultaneously supported in the system. 
\appendices
\vspace{-7pt}
\section{AMP based Inner Decoder}\label{sec:appendix}

In the following we present the details of the inner decoder. We describe the inference procedure for the extended system model with 
a multi-antenna receiver in a block-fading scenario\footnote{The derivations are given here for the \ac{MMV-AMP} algorithm for the multi-antenna system model. The \ac{AMP} algorithm for the single-antenna system model follows naturally from the \ac{MMV-AMP} by setting $T=1$.}. The task of the inner decoder consists of detecting the transmitted sequences and is realized for both communication procedures via the same algorithmic solution as described in Section~\ref{sec:decoding}, i.e., the \ac{MMV-AMP} algorithm \cite{kim2011belief}. 

Let $c_k\in\{0,1\}$ denote the binary random variable which indicates if sequence $k$ has been sent with  probability $P(c_k=1)=\epsilon_k$ and $\mathbf{h}_k\in\mathbb{C}^T$ the corresponding channel vector of the $k$th sequence. \textcolor{black}{The probability $\epsilon_k$ should be initialized such that the sparsity pattern of $\mathbf{X}$ is reflected. Given that each user employs a CW code construction $\mathcal{C}_{\mathrm{CW}}=\mathrm{PPM}(q)\circ \mathrm{RS}(n',k')$, the number of active users $\Ka$ and the total number of users denoted by $\Ktot$, $\epsilon_k$ for S-RA is given by $\epsilon_k^{\mathrm{(S)}}= \frac{\Ka}{q\Ktot}$ and for U-RA by $\epsilon_k^{\mathrm{(U)}}= \frac{\Ka}{q}$.} The row-wise distribution of $\mathbf{x}_k\triangleq c_k\mathbf{h}_k$ is given by
\begin{equation}
    P_{\mathbf{x}_k}(\mathbf{x}_k)=(1-\epsilon_k)\delta(\mathbf{h}_k)+\epsilon_k P_{\mathbf{h}_k}(\mathbf{h}_k)\label{eq:prior},
\end{equation}
where $P_{\mathbf{h}_k}(\mathbf{h}_k)$ denotes the channel distribution, where $\mathbf{h}_k\sim\mathcal{CN}(0,\mathbf{I})$ (Rayleigh fading). The iterations of the MMV-AMP algorithm are summarized in Algorithm~\ref{alg:inner_decoder}. The non-linear denoising function $\eta:\mathbb{C}^{N\times T}\times \mathbb{R}_{+}^T\times \{0,1\}^N\rightarrow \mathbb{C}^{N\times T}$ in line \ref{eq:denoiser} is defined as
\begin{equation}
    \eta(\mathbf{R},\boldsymbol{\tau},\boldsymbol{\epsilon}):=[\eta_1(\mathbf{r}_1,\boldsymbol{\tau}_1),\ldots,\eta_N(\mathbf{r}_{N},\boldsymbol{\tau}_N)]^{\mathrm{T}}\label{eq:denoising},
\end{equation}
where the row-wise denoising function $\eta_k:\mathbb{C}^T\times \mathbb{R}_{+}^T\times\{0,1\}$ computes the posterior mean estimate of the random vector $\mathbf{x}_k$, with prior distribution defined in \eqref{eq:prior} and the decoupled Gaussian likelihood assumption $\mathbf{r}_k=\mathbf{x}_k+\mathbf{z}_k$, such that, $\mathbf{z}_k\sim\mathcal{CN}(\boldsymbol{0},\mathrm{diag}(\boldsymbol{\tau}))$. Given the posterior mean defined by $\eta_k(\mathbf{r}_k,\boldsymbol{\tau},\epsilon_k):=\mathbb{E}[\mathbf{x}_k\vert \mathbf{r}_k,\boldsymbol{\tau}]$, for Rayleigh fading \eqref{eq:denoising} is given by
\begin{equation}
\eta_k(\mathbf{r}_k,\boldsymbol{\tau},\epsilon_k)=\phi(\mathbf{r}_k)(\mathbf{I}+\boldsymbol{\Sigma}_k)^{-1}\mathbf{r}_k,
\end{equation}
with $\phi(\mathbf{r}_k)=(1+D_k\frac{(1-\epsilon_k)}{\epsilon_k} \exp(-\mathbf{r}_k^{H}(\boldsymbol{\Sigma}_k^{-1}+\tilde{\boldsymbol{\Sigma}}_k^{-1})\mathbf{r}_k))^{-1}$, $\boldsymbol{\Sigma}_k=\mathrm{diag}(\boldsymbol{\tau}_k)$, $\tilde{\boldsymbol{\Sigma}}_k=\mathbf{I}+\boldsymbol{\Sigma}_k$ and $D_k=\vert\boldsymbol{\Sigma}_k+\mathbf{I}\vert/\vert\boldsymbol{\Sigma}_k\vert$. The term $\eta'$ in step \ref{eq:eta_prime} is defined for the Rayleigh fading by the Jacobi matrix
\begin{equation}
    \eta'(\mathbf{r}_k,\boldsymbol{\tau},\epsilon_k)=\phi(\mathbf{r}_k)(\mathbf{I}+\Sigma_k)^{-1}+(\tilde{\boldsymbol{\Sigma}}_k^{-1}\mathbf{r}_k)(\tilde{\boldsymbol{\Sigma}}_k^{-1}\boldsymbol{\Sigma}_k^{-1}\mathbf{r}_k)^H(\phi(\mathbf{r}_k)-\phi(\mathbf{r}_k)^2).
\end{equation}
\smallskip
\vspace{-7pt}
\begin{minipage}[t!]{0.479\textwidth}
{
\begin{algorithm}[H]
\caption{Inner decoder}\label{alg:inner_decoder}
\begin{algorithmic}[1]
\algsetup{linenosize=\small}
\scriptsize
\renewcommand{\algorithmicrequire}{\textbf{Input:}}
\renewcommand{\algorithmicensure}{\textbf{Initialize:}}
\REQUIRE $\mathbf{Y}$, $\mathbf{A}$, $\boldsymbol{\epsilon}$, $\theta$
\ENSURE $\mathbf{X}^0=0,\;\mathbf{Z}^0=\mathbf{Y}$
\FOR {$t=1$ to $T_{\mathrm{Max}}$} 
\STATE $\boldsymbol{\tau}_{i}^{(t)}=\theta\Vert\mathbf{Z}^{(t-1)}\Vert^2_2 +(1-\theta)\boldsymbol{\tau}_{i}^{(t-1)}$
\STATE $\mathbf{X}^{(t)}=\eta(\mathbf{A}^{\mathrm{H}}\mathbf{Z}^{(t-1)}+\mathbf{X}^{(t-1)},\boldsymbol{\tau}^{(t)},\boldsymbol{\epsilon})$\label{eq:denoiser}
\STATE $\mathbf{R}^{(t)}=\theta(\mathbf{A}^{\mathrm{H}}\mathbf{Z}^{(t)}+\mathbf{X}^{(t)})+(1-\theta)\mathbf{R}^{(t-1)}$
\STATE $\hat{\mathbf{Z}}=\mathbf{A}\mathbf{X}^{(t)}$
\STATE $\mathbf{Z}^{(t)}=\mathbf{Y}-\hat{\mathbf{Z}}+\frac{N}{n}\mathbf{Z}^{(t-1)}\langle\eta'(\mathbf{R}^{(t)},\boldsymbol{\tau}^{(t)})\rangle$ \label{eq:eta_prime}
\ENDFOR
\end{algorithmic}
\end{algorithm}
}
\end{minipage}
\hfill
\begin{minipage}[t!]{0.48\textwidth}
{
\begin{algorithm}[H]
\caption{Detection for S-RA.}\label{alg:sourced_decoder}
\begin{algorithmic}[1]
\algsetup{linenosize=\small}
\scriptsize
\renewcommand{\algorithmicrequire}{\textbf{Input:}}
\renewcommand{\algorithmicensure}{\textbf{Initialize:}}
\REQUIRE $\mathbf{Y}$, $\ASD$, $\rho_l$
\ENSURE $\epsilon_k(0)=\frac{\rho_l}{M}$
\FOR {$i=1$ to $L_{\mathrm{Max}}$} 
\STATE $\mathbf{R},\boldsymbol{\tau}_{i}^2\rightarrow \mathrm{Algorithm} \;\ref{alg:inner_decoder}$
\STATE $\hat{\rho}_l(i)\rightarrow \eqref{eq:activity_update}$
\STATE $\hat{\boldsymbol{\epsilon}}_{l}(i)\rightarrow\eqref{eq:message_update}$
\STATE $\boldsymbol{\epsilon}(i+1)\rightarrow\text{SISO}(\hat{\boldsymbol{\epsilon}}(i))$\cite{bimberg10}
\ENDFOR
\end{algorithmic}
\end{algorithm}
}
\end{minipage}

\section{Iterative Bayesian Detection for S-RA}\label{sec:AD}
The S-RA setup allows for an iterative Bayesian detection procedure in which beliefs are iteratively passed between an inner decoder, i.e., Algorithm \ref{alg:inner_decoder}, and an outer decoder, realized by the \ac{SISO} decoder for non-binary codes from \cite{bimberg10}. 
In the following we derive the user activity detection step and the sparsity update and summarize the overall algorithm in Algorithm~\ref{alg:sourced_decoder}. For the beliefs that are passed to the outer decoder we have
\begin{equation}
    \hat{\epsilon}_k= (1+\exp-\mathrm{LLR}_{k})^{-1}
\end{equation}
with $\mathrm{LLR}_{k}=\mathbf{r}_k^{H}\boldsymbol{\Sigma}_k\tilde{\boldsymbol{\Sigma}}_k\mathbf{r}_k-\log D_k$.
Assume that every active user $l\in[K]$ uses the CW construction $\mathcal{C}_{\mathrm{CW}}=\mathrm{PPM}(q)\circ \mathrm{RS}(n',k')$ to encode its message into a binary message vector $\mathbf{c}_l=[\mathbf{c}_l^{(1)};\ldots;\mathbf{c}_l^{(n')}]$, where we denote by $\mathbf{c}_l^{(i)}$ the $i$-th PPM encoded symbol of an $\mathrm{RS}(n',k')$ codeword. 
Let the pdf of the CW code $\mathbf{c}_l$ of the $l$th user be
\begin{equation}
    P(\mathbf{c}_l\vert\lambda_l)=\prod_{i=1}^{n'} P(\mathbf{c}_l^{(i)}\vert\lambda_l)\label{eq:prior41},
\end{equation}
where for every sub-vector $\mathbf{c}_l^{(i)}$ we have
\begin{equation}
    P(\mathbf{c}_l^{(i)}\vert\lambda_l)=\begin{cases}
        \frac{1}{q}\sum\limits_{j=1}^{q}\delta(c_{jl}^{(i)}-1)\prod\limits_{k\neq j}^{M}\delta(c_{kl}^{(i)})&\mathrm{for}\; \lambda_l=1\\
        \prod\limits_{j=1}^{q}\delta(c_j) &\mathrm{for}\;\lambda_l=0.
    \end{cases},\label{eq:prior4}
\end{equation}
with $c_{kl}^{(i)}$ denoting the $k$-th element of $\mathbf{c}_l^{(i)}$. 
We marginalize \eqref{eq:prior41} with respect to $\mathbf{c}_l$, i.e., 
\begin{equation}
    \nu^{(i)}(\lambda_l)\propto\begin{cases}
        \frac{1}{q}\sum_j\epsilon_{jl}^{(i)}\prod_{i\neq j}(1-\epsilon_{il}^{(i)}),&\lambda_l=1\\
        \prod_{j}(1-\epsilon_{jl}^{(i)}), & \lambda_l=0
    \end{cases},
\end{equation}
where we denote by $\nu^{(i)}$ the partial belief from marginalizing with respect to $\mathbf{c}_l^{(i)}$. The resulting partial beliefs can be formulated in LLR form by
\begin{equation}
    \mathrm{LLR}_{\lambda}^{(i)}=\log \sum_{j}\epsilon_{jl}^{(i)}/(1-\epsilon_{jl}^{(i)})-\log q,
\end{equation}
from which the LLRs for the user activity update can be given by
\begin{equation}
    \mathrm{LLR}_{\lambda}=\sum_i\mathrm{LLR}_{\lambda}^{(i)}+\log \rho_l/(1-\rho_l),
\end{equation}
and the respective belief updates by
\begin{equation}
    \hat{\rho}_l\triangleq (1+\exp -\mathrm{LLR}_{\lambda})^{-1}\label{eq:activity_update}.
\end{equation}
Given \eqref{eq:activity_update}, the belief update for every sequence can now be stated as
\begin{equation}    
    \nu_{f_{\mathbf{c}}\rightarrow c}(c_{jl}^{(i)},\lambda_l)\propto    \sum_{\mathbf{c}/c_{jl}^{(i)}}P(\mathbf{c}_l^{(i)}\vert\lambda_l)\prod_{r\neq j}\nu_{c\rightarrow f_{\mathbf{c}}}(c_{rl}^{(i)}),
\end{equation}
for which after marginalization we get
\begin{equation}
    \nu_{f_{\mathbf{c}}\rightarrow c}(c_{jl}^{(i)})\propto\begin{cases}
    \frac{\hat{\rho}_l}{q}\epsilon_{jl}^{(i)}\prod\limits_{r\neq j}(1-\epsilon_{rl}^{(i)}),& c_j=1\\
    \hat{\rho}_l\sum\limits_{r\neq j}\epsilon_{rl}^{(i)}\prod\limits_{k\neq r}(1-\epsilon_{kl}^{(i)})+(1-\hat{\rho}_l)\prod\limits_{k}(1-\epsilon_{kl}^{(i)}),& c_j = 0
    \end{cases}.
\end{equation}
As result, the updated beliefs for every sequence are given by
\begin{equation}    
\hat{\epsilon}_{jl}^{(i)}=(1+\frac{(1-\epsilon_{jl}^{(i)})}{\epsilon_{jl}^{(i)}}(\sum\limits_{r\neq j}\frac{\epsilon_{rl}^{(i)}}{(1-\epsilon_{rl}^{(i)})}+\frac{(1-\hat{\rho}_l)}{\hat{\rho}_l}))^{-1}\label{eq:message_update},
\end{equation}
which are passed as likelihoods to the outer decoder.
\vspace{-7pt}
\section{Details on Constant Weight Decoding}\label{sec:appendix_outer_decoder}

The task of the outer decoder is to retrieve the binary codewords of the users that are active over the resource block, $\mathbf{c}_{i_1}, \ldots,\mathbf{c}_{i_{\Ka}}$, from the output of the inner decoder. The decoding is based on the procedure in \cite{AbdulJabbar88} for decoding of superimposed codes over multiaccess OR channels. There, the task is to map the binary superimposed sequence $\mathrm{f}(\mathcal{A})$ formed by the multiaccess OR channel into a set of codewords $\mathcal{A}=\{c_i\in\mathcal{C},\: i=1, \ldots, K\}$ from a given superimposed code $\mathcal{C}$. 
Let $\mathrm{f}(\mathcal{A})$ be the superposition of a set $\mathcal{A}$ of codewords from a CW code $\mathcal{C}\in\mathcal{CW}\left(n,m,w,d\right)$ obtained by concatenating a $\mathrm{PPM}(q)$ code with a systematic $\mathrm{RS}(n,k)$ code. The binary sequence $\mathrm{f}(\mathcal{A})$ can be split into $n$ consecutive sub-vectors denoted by $\mathbf{f}_n=[f_{0n},\ldots,f_{(q-1)n}]$.
Let $S_j(\mathrm{f}(\mathcal{A}))$ denote the set of $\mathbb{GF}(q)$ symbols defined by the mapping
\begin{equation}
    S_j(\mathrm{f}(\mathcal{A}))=\left\{i\in\mathbb{GF}(q)\mid\mathrm{f}_{ij}=1\right\},\quad j\in[n],
\end{equation}
and $\mathbf{G}_{RS}$ denote the generator matrix of a systematic RS code over $\mathbb{GF}(q)$. Let further $B(\mathrm{f}(\mathcal{A}))$ denote a set of codewords from the code $\mathcal{C}$ defined by
\begin{equation}
    B(\mathrm{f}(\mathcal{A}))=\left\{\mathbf{F}\left(\mathrm{G}_{RS}\cdot\mathbf{i}\right)|\mathbf{i}=(\mathrm{i}_1,\mathrm{i}_2,\dots,\mathrm{i}_k), \,\,\mathrm{i}_l\in S_j(\mathrm{f}(\mathcal{A}))\right\},
\end{equation}
where $\mathbf{F}(\mathbf{c}_{RS})$ denotes the transformation from the RS codeword into a PPM(q) codeword. The decoder for the outer code can then be defined as the search procedure that produces following set of codewords
\begin{equation}
    \hat{\mathcal{A}}=\left\{\mathbf{c}\in B(\mathrm{f}(\mathcal{A}))\mid \langle\mathbf{c},\mathrm{f}(\mathcal{A})\rangle=w\right\}.
\end{equation}
The decoding complexity of the search decoder is equal to $\mathcal{O}(\Ka^k)$.
%
\ifCLASSOPTIONcaptionsoff
\newpage
\fi
%
\bibliographystyle{IEEEtran}
\bibliography{IEEEabrv,./bibliography/bibliography}
\end{document}